\documentclass[final,5p,times,twocolumn,authoryear]{elsarticle}

\journal{ } 

\usepackage{epsfig}
\usepackage{amssymb,amsmath,amsthm,amsfonts}
\usepackage[T1]{fontenc}
\usepackage{bm}
\usepackage{mathrsfs}
\usepackage{listings} 
\usepackage{float}
\usepackage{relsize}
\usepackage{booktabs}
\usepackage{enumitem}
\usepackage{multirow}
\usepackage[toc,page]{appendix}
\usepackage{mathdots}
\usepackage{graphicx}
\usepackage{graphics}
\usepackage{caption}
\usepackage{subcaption}
\numberwithin{equation}{section}
\usepackage{arydshln}
\usepackage{xcolor}
 
\newcommand{\N}{{\mathord{\mathbb N}}}
\newcommand{\Prob}{{\mathord{\mathbb P}}}
\newcommand{\normal}{{\mathcal{ N}}}

\newcommand{\R}{{\mathord{\mathbb R}}}

\newcommand{\trans}{{\mathbf A}}
\newcommand{\emission}{{\mathbf B}}

\newcommand{\emissionmu}{{\bm \mu}}
\newcommand{\emissionsigma}{{\bm \Sigma}}
\newcommand{\transpi}{{\bm \pi}}

\newcommand{\E}{{\mathord{\mathbb E}}}

\newcommand{\figscale}{1.0}

\begin{document}


\begin{frontmatter}

\title{Predicting Risk-adjusted Returns using an Asset Independent Regime-switching Model}

\author[sorbonne]{Nicklas Werge\corref{c1}}
\ead{nicklas.werge@upmc.fr}
\address[sorbonne]{LPSM, Sorbonne Universit\'e, 4 place Jussieu, 75005 Paris, France}
\cortext[c1]{Corresponding author}

\begin{abstract}
Financial markets tend to switch between various market regimes over time, making stationarity-based models unsustainable.
We construct a regime-switching model independent of asset classes for risk-adjusted return predictions based on hidden Markov models.
This framework can distinguish between market regimes in a wide range of financial markets such as the commodity, currency, stock, and fixed income market.
The proposed method employs sticky features that directly affect the regime stickiness and thereby changing turnover levels.
An investigation of our metric for risk-adjusted return predictions is conducted by analyzing daily financial market changes for almost twenty years.
Empirical demonstrations of out-of-sample observations obtain an accurate detection of bull, bear, and high volatility periods, improving risk-adjusted returns while keeping a preferable turnover level.
\end{abstract}

\begin{keyword}
hidden Markov model \sep financial time series \sep non-stationary \sep regime-switching \sep prediction markets \sep trading strategies
\end{keyword}
\end{frontmatter}


\section{Introduction} \label{sec:introduction}


Financial markets are known to shift between economic cycles; some of the most well-known regimes are the bull, bear, and high-volatility markets.
Each of these market regimes may have financial characteristics unique to this particular regime.
One of the most common methods of financial market analysis is time series analysis.
Time series models are used to predict future prices,  price changes, and volatilities in a wide range of financial markets.
Some of the most famous models are the AutoRegressive Integrated Moving Average (ARIMA) models.
However, analyzing financial time series through these traditional time series methods may result in misleading resolutions as they cannot embrace the nonlinear characteristics of financial time series, e.g., the stationarity assumption often seems dubious in practice.
Therefore, non-stationary-based time series models are more suitable for financial time series.
One could comprehend this by modifying these time series models by incorporating a time-dependent variable to adjust for the non-stationarity, e.g., the threshold autoregressive time series model.

Another way to capture financial markets' tendency to switch between regimes is the Hidden Markov Model (HMM), as it "only" assumes local or state-conditioned stationarity.
Modeling times series data using HMMs became mainstream after \cite{baum1970} and \cite{rabiner1989} applied it across many areas (e.g., speech recognition, medical applications, and text classification).
The idea of making a Markov-switching approach to analyze financial time series became popular after \cite{hamilton1989} applied this approach to identify economic cycles of GNP levels.
More recently, the HMM has been used to predict market regimes in the financial markets due to their ability to capture multiple characteristics from financial return series such as time-varying correlations, fat tails, volatility clustering, skewness, and kurtosis, while also providing reasonable approximations even for processes in which the underlying model is unknown (\cite{ang2012,nystrup2015,nystrup2017}).
Besides, HMMs are advantageous as they allow ample interpretability of the results; thinking in market regimes is a natural approach for financial practitioners.
Nevertheless, the lack of data availability makes the linking between investment purposes and business cycles a complex and challenging task.
As the market regimes are not observable, one has to extract them from the time series.
However, this extraction is not unambiguous, as some specific regimes may be up for discussion in the financial practitioner's community, e.g., high and low volatility regimes depend on the given risk-aversion.
Consequently, we demand a model to apprehend the various economic sentiments of the financial markets.

Many researchers have applied HMMs to analyze and predict economic (non-linear) trends and future financial asset prices.
\cite{kritzman2012} studied an HMM with two states to predict regimes in market turbulence, inflation, and economic growth index.
\cite{hassan2005} and  \cite{nguyen2018} used the HMM to forecast prices in the stock market.
A combination of open, close, low, and high prices was used in \cite{gupta2012} for stock price prediction.
All of the above references use four hidden states in their study on the stock market.
\cite{guidolin2007a} and \cite{fons2019} used a four-state and two-state HMM, respectively, in their studies of asset allocation decisions using various time series.
As suggested by \cite{guidolin2007b}, a range between two and four hidden states in the HMM is often encountered in financial studies.
However, studies of applying HMMs to predict trends across a broad range of assets are sparse.

In this study, we focus on predicting risk-adjusted returns using a single regime-switching model. 
Using only one HMM to analyze a wide range of assets, we enforce generalizations in the model.
This framework is made with so-called "sticky" features that naturally enhance regime stickiness by an adjustable hyperparameter.
Finally, we demonstrate our methodology on a broad range of asset classes by analyzing daily financial market changes for almost twenty years. 
The investigation illustrates our metric ability to predict risk-adjusted returns for different regime stickiness choices.
Our experiments are conducted using out-of-sample observations, showing an accurate detection of bull, bear, and high volatility periods, improving risk-adjusted returns while keeping a preferable turnover level.


\section{Hidden Markov Models (HMMs)} \label{sec:hmm}


There is much literature about HMMs, but to have the necessary notions, we briefly sketch the elements of the HMM, how to estimate the parameters, and select the number of hidden states in the HMMs.
For a comprehensive introduction of the inference of HMMs, we refer to \cite{zucchinimacdonald2009} and \cite{murphy2013}.


\subsection{Elements of HMM} \label{sec:hmm_elements}


The HMM is a probabilistic model in which a sequence of observations $x=(x_1,\dots, x_n)$ with $x_t \in \R^d$ for $t=1, \dots, n$ is generated by a latent finite-state Markov chain $z=(z_1, \dots , z_n)$.
Denote by $d$ the dimension of the observations.
We call $z$ the sequence of hidden states where $z_t \in \{ 1, \dots, S \}$ for $t=1, \dots, n$ with $S$ the number of hidden states.
The HMM can be specified by the initial probability vector $\transpi = \{ \pi_{i} \}_{i=1, \dots S} \in \R^{S}$, a transition probability matrix $\trans = \{ A_{ij} \}_{i,j=1,\dots,S} \in \R^{S \times S}$ and the emission probabilities $\emission$ which can be any distribution conditioned on the current hidden state. 
The parameters of the HMM are given by $\Lambda = \{\transpi, \trans, \emission \}$ and have to be estimated from the observed sequence $x$. 
Note that $\pi_{i}=\Prob(z_1 = i)$ is the probability for being in hidden state $i$ at time $t=1$ where $\sum_{i=1}^{S} \pi_i =1$, $A_{ij} = \Prob(z_{t}=j \vert z_{t-1}=i)$ is the transition probability of moving from hidden state $i$ at time $t-1$ to hidden state $j$ at time $t$ with $\sum_{j=1}^{S} A_{ij} = 1$, and $\emission$ is the parameters of the conditional densities $p(x_t \vert z_t=j)$.

When working with financial time series, a typical choice of emission probabilities is the Gaussian Mixture Model (GMM).
However, other density functions could likewise be considered.
A gentle introduction of HMMs with GMM emissions is made in \cite{bilmes1998}.
The authors of \cite{ang2012} and \cite{nystrup2015}  show evidence on the HMMs ability to comprehend several stylized facts, such as leptokurtosis, heteroskedasticity, skewness, and time-varying correlations, by use of the GMM as emission probability.
For simplicity, we assume the distribution of emission probabilities $\emission$ to be Gaussian; $\emission = p(x_t \vert z_t = j, \Lambda) = \normal(x_t\vert \mu_j, \Sigma_j)$ where $\emissionmu = \{ \mu_j \}_{j=1,\dots ,S}$  is the mean vectors and $\emissionsigma = \{ \Sigma_j \}_{j=1,\dots ,S}$ the co-variance matrices with $\mu_j \in \R^{d}$ and $\Sigma_j \in \R^{d \times d}$ for $j=1,\dots, S$. 
Thus, the model parameters of our HMM is given as $\Lambda = \{\transpi, \trans, \emissionmu, \emissionsigma \}$.


\subsection{Parameter Estimation} \label{sec:hmm_estimation}


There are three fundamental problems in estimating the HMM:
\begin{itemize}
\item Given the observations sequence $x=(x_1,\dots, x_n)$ and HMM parameters $\Lambda = \{\transpi, \trans, \emissionmu, \emissionsigma \}$, how can we estimate $\Prob (x | \Lambda )$ the likelihood of the given observation sequence.
\item Given the observations sequence $x=(x_1,\dots, x_n)$ and model parameters $\Lambda = \{\transpi, \trans, \emissionmu, \emissionsigma \}$, how can we choose a sequence of hidden states $z=(z_1, \dots , z_n)$, which is optimal.
\item How do we adjust the HMM parameters $\Lambda = \{\transpi, \trans, \emissionmu, \emissionsigma \}$ to maximize $\Prob (x | \Lambda )$.
\end{itemize}
There are several approaches to solve these problems since there are several possible optimal criteria.
We choose to solve the first and the second problem by the dynamic programming algorithms known as the forward-backward algorithm proposed by \cite{baum1967} and \cite{baum1968}, and the Viterbi algorithm (\cite{viterbi1967}).
The third problem is solved by the iterative Baum-Welch (BW) algorithm, a type of the Expectation-Maximization (EM) algorithm (\cite{rabiner1989}).

The BW algorithm alternates between an expectation step and a maximization step until convergence is reached, often abbreviated as the E-step and M-step.
In the E-step, we calculate the expected log-likelihood of the hidden state given the observation sequence $x$ and model parameters $\Lambda$.
Next, in the M-step we maximize the expected log-likelihood from the E-step to update our model parameters $\Lambda$.
We denote by $Q(\Lambda, \bar{\Lambda})$ the function of the expectation of the complete log-likelihood given as
\begin{align} \label{eq:q_function}
Q(\Lambda, \bar{\Lambda}) = \E [ \log \Prob( x, z | \Lambda)  | x, \bar{\Lambda}),
\end{align}
where the current model is $\Lambda$ and the previous model as $\bar{\Lambda}$. 

It can be proven that $\Prob ( x | \Lambda) \geq \Prob ( x | \bar{\Lambda})$, but it is essential to remember that the BW algorithm does not guarantee a global solution. 
As suggested in \cite{adams2016} and \cite{fons2019}, we modify the $Q(\Lambda, \bar{\Lambda})$ function with the priors of the model parameters $G(\Lambda)$, namely
\begin{align} \label{eq:map_function}
Q(\Lambda, \bar{\Lambda}) + \log(G(\Lambda)),
\end{align}
which is called Maximum a Posteriori (MAP) estimation (\cite{gauvain1994}).
Thus, in the E-step, we calculate the $Q(\Lambda, \bar{\Lambda})$ function from \eqref{eq:q_function}, and for the M-step, we maximize \eqref{eq:map_function}.


\subsection{Prediction} \label{sec:hmm_prediction}


The prediction of the hidden states sequence $(z_1, \dots, z_n)$ is estimated using the observation sequence $(x_1, \dots, x_n)$ as described in Section \ref{sec:hmm_estimation}.
We denote by $\alpha_{n \vert n}$ the vector of state probabilities at time $n$ (given the sequence of observations $x=(x_1,\dots, x_n)$) with the $j$th entry $(\alpha_{n \vert n})_j = \Prob(z_n=j \vert x)$ for $j=1,\dots, S$. 
Thus, one can forecast the state probability $h \geq 0$ steps ahead by \begin{align} \label{eq:pred_hs_sequence}
\alpha_{n+h \vert n} = \alpha_{n \vert n} \trans^{h},
\end{align}
as the model parameters $\trans$ are assumed to be constant over time.


\subsection{Model Selection} \label{sec:hmm_modelseletion}


A drawback of using the HMM is the necessity of knowing the number of hidden states in advance (such as the hyper-parameter $k$ in the $k$-nearest neighbor algorithm and $k$-means clustering). 
There are several criteria used for this model selection: the lazy approach is to use statistical criteria such as the Akaike's Information Criterion (AIC) by \cite{akaike1974}, Bayesian Information Criterion (BIC) by \cite{schwarz1978}, Hannan-Quinn Information Criterion (HQIC) by \cite{hannan1979}, and Bozdogan Consistent Akaike Information Criterion (BCAIC) by \cite{bozdogan1987}. 
These criteria are defined as follows:
\begin{align*}
\text{AIC} &= -2 \log(L) + 2p ,
\\ \text{BIC} &= -2 \log(L) + p \log(n),
\\ \text{HQIC} &= -2 \log(L)  + p \log(\log(n)),
\\ \text{BCAIC} &= -2 \log(L)  + p (\log(n)+1),
\end{align*}
where $\log(L)$ is the log-likelihood of the model, $n$ indicates the number of observations in the time series, and $p$ denotes the number of independent parameters of the model. 
In the case of an HMM with GMM emissions, we have $p = S(S + cm)$, where $S$ is the number of hidden states in the Markov chain of the model, $m$ is the number of Gaussian mixtures, and $c$ is the number of parameters of the underlying distribution of the observation process. 
Note that a $d$-dimensional multivariate Gaussian with full covariance matrix process has $c = d+d(d+1)/2$ parameters to estimate. 
Thus, an HMM with three hidden states $(S=3)$, a single $2$-dimensional Gaussian process in each hidden state, has a total of $24$ parameters.

Suppose one were to see the number of hidden states as the number of strategies we have to make to produce proper predictions.
Then the number should be neither too small nor too large.
If the number of hidden states is too small, then the risk of misclassification will increase.
Too many hidden states will make the distinction between each hidden state vague and, therefore, increase the risk for overfitting and increase the computational cost.
A similar observation can be made regarding the number of Gaussian mixture components.

However, if one wishes to maintain a high degree of interpretability of the hidden states in the model, we should keep the number of hidden states low.
Another approach is the greedy approach, where we decide the number of hidden states in the HMM by constructing different portfolios based on HMMs with different numbers of hidden states and then select the number of hidden states associated with the portfolios of the best performance, e.g., evaluated by the Sharpe Ratio (SR).
One should be aware that we may find different optimal numbers of states for each asset using these criteria. 


\section{Data} \label{sec:data}


Our objective is to identify market regimes on various asset classes, namely commodity (CO), currency (FX), equity (EQ), and fixed income (FI). 
We consider $d=15$ instruments defined as $I = (I_{1}, \dots, I_{15})^T$, consisting of four different instruments per asset type, except for commodities where we have only three instruments.
All instruments $I$ are future contracts generated automatically by selecting the nearest contract. 
The data analyzed are closing returns of daily frequency from January 2000 to October 2019, consisting of $n=4972$ observations (per instrument).

Table \ref{tab:data_perf} presents an overview of the performance of each asset.
This confirms a high degree of variation of the considered asset; commodities and equities are the most volatile asset classes, whereas fixed income volatility is several times lower. 
Currencies appear to be in the middle of the levels we observe for equities and fixed income.
Fixed income seems to be the most coherent asset class, whereas we find some large variations in returns, volatility, and maximum drawdown in commodities.

\begin{table}[!htbp]
\centering
\begin{tabular}{clrrrr}
\hline
\textbf{\#} & \textbf{Instr.} & \textbf{Ret.} & \textbf{Vol.} & \textbf{SR} & \textbf{DD} \\ \hline
1 & CO1 & $ 6.36 \% $ & $ 17.23 \% $ & $ 0.45 $ & $ 16.81 \% $ \\
2 & CO2 & $ -20.19 \% $ & $ 50.4 \% $ & $ -0.2 $ & $ 37.71 \% $ \\
3 & CO3 & $ 5.46 \% $ & $ 36.12 \% $ & $ 0.33 $ & $ 24.13 \% $ \\ \hdashline
4 & FX1 & $ -0.83 \% $ & $ 9.17 \% $ & $ -0.04 $ & $ 10.83 \% $ \\
5 & FX2 & $ -0.02 \% $ & $ 9.58 \% $ & $ 0.05 $ & $ 5.83 \% $ \\
6 & FX3 & $ -1.82 \% $ & $ 9.79 \% $ & $ -0.14 $ & $ 8.29 \% $ \\
7 & FX4 & $ -0.98 \% $ & $ 7.96 \% $ & $ -0.08 $ & $ 4.98 \% $ \\
\hdashline
8 & EQ1 & $ 1.1 \% $ & $ 23.34 \% $ & $ 0.16 $ & $ 18.39 \% $ \\
9 & EQ2 & $ 2.64 \% $ & $ 18.1 \% $ & $ 0.24 $ & $ 15.79 \% $ \\
10 & EQ3 & $ 1.08 \% $ & $ 24.61 \% $ & $ 0.17 $ & $ 27.98 \% $ \\\
11 & EQ4 & $ 4.14 \% $ & $ 18.67 \% $ & $ 0.31 $ & $ 21.02 \% $ \\
\hdashline
12 & FI1 & $ 3.73 \% $ & $ 5.91 \% $ & $ 0.66 $ & $ 5.02 \% $ \\
13 & FI2 & $ 4.22 \% $ & $ 5.24 \% $ & $ 0.82 $ & $ 3.49 \% $ \\
14 & FI3 & $ 3.65 \% $ & $ 5.88 \% $ & $ 0.65 $ & $ 4.19 \% $ \\
15 & FI4 & $ 2.35 \% $ & $ 3.05 \% $ & $ 0.79 $ & $ 2.87 \% $ \\
\hline
\end{tabular}
\caption{Performance overview (annualized return, annualized volatility, Sharpe ratio, and maximum drawdown) of our instruments $I$ evaluated from January 2000 to October 2019. Note $I \in \R^{n \times d}$ with $n=4972$ and $d=15$.}
\label{tab:data_perf}
\end{table}

To further emphasize our instruments' diversity, we show the range (minimum; maximum) of the one-year rolling mean, standard deviation, skewness, and (excess) kurtosis in Table \ref{tab:statistics}.
The instruments $I$ show a considerable amount of variability, both within and across instrument types, with commodities showing the most variation and fixed income showing the least fluctuation.
In particular, it is not abnormal that skewness exceeds one (in absolute terms), nor kurtosis is negative (platykurtic) or very positive (leptokurtic), e.g., CO2 have a kurtosis above thirty-five.

\begin{table}[!htbp]
\centering
\setlength{\tabcolsep}{2pt}
\begin{tabular}{clllll}
\hline
\textbf{\#} & \textbf{Instr.} & \textbf{Mean} & \textbf{Std.} & \textbf{Skew.} & \textbf{Kurt.} \\ 
\hline
1 & CO1 & (-3.26;3.93) & (5.09;23.69) & (-3.63;2.20) & (-0.42;19.33) \\
2 & CO2 & (-8.94;12.6) & (14.01;58.96) & (-1.01;4.15) & (-0.64;35.64) \\
3 & CO3 & (-13.48;5.33) & (8.61;56.73) & (-2.54;1.79) & (-0.69;11.50) \\ 
\hdashline
4 & FX1 & (-3.12;1.6) & (2.7;13.02) & (-4.18;0.97) & (-0.66;25.85) \\
5 & FX2 & (-2.37;1.57) & (2.35;12.07) & (-1.36;1.10) & (-0.61;6.59) \\
6 & FX3 & (-1.89;2.12) & (2.67;11.58) & (-2.29;2.65) & (-0.36;13.96) \\
7 & FX4 & (-1.35;1.92) & (2.07;9.87) & (-0.88;1.30) & (-0.70;4.62) \\ 
\hdashline
8 & EQ1 & (-5.55;2.61) & (5.84;35.71) & (-2.01;1.90) & (-0.42;13.29) \\
9 & EQ2 & (-5.03;2.41) & (4.66;32.84) & (-1.41;1.21) & (-0.56;5.26) \\
10 & EQ3 & (-8.62;4.55) & (6.47;49.24) & (-2.61;1.25) & (-0.80;13.03) \\
11 & EQ4 & (-5.77;2.16) & (3.64;39.59) & (-4.05;1.49) & (-0.43;22.06) \\ 
\hdashline
12 & FI1 & (-0.86;1.8) & (1.66;8.11) & (-1.40;1.39) & (-0.75;6.52) \\
13 & FI2 & (-0.62;1.34) & (1.97;6.25) & (-1.32;0.70) & (-0.77;3.83) \\
14 & FI3 & (-0.81;1.4) & (1.94;7.3) & (-0.90;1.60) & (-0.70;6.92) \\
15 & FI4 & (-0.74;0.58) & (0.48;4.49) & (-4.34;2.24) & (-0.49;26.33) \\
\hline
\end{tabular}
\caption{Range $(\min;\max)$ of one-year rolling mean, standard deviation, skewness and (excess) kurtosis of instruments $I$. 
Rolling mean and standard deviation are scaled by $10^{3}$.}
\label{tab:statistics}
\end{table}


\section{Feature Engineering} \label{sec:feature_eng}



\subsection{Exponential Weighted Moving Moments} \label{sec:ewmm}


When the underlying parameters are believed to follow a random walk, it is natural to use exponential forgetting.
One of the most popular methods for calculating moments is the Exponential Weighted Moving Moment (EWMM) method, which is applied extensively in many different fields due to its computational efficiency.
This EWMM method is often used to reduce noisy time-series data, also called "smoothing" the data.
We can define the $\text{EWMM}^{i}_{t}$ of order $i \in \N$ at time $t$ by
\begin{align*}
\text{EWMM}^{i}_{t} = \lambda \text{M}^{i}_{t} + (1-\lambda)\text{EWMM}^{i}_{t-1},
\end{align*}
where $\lambda = \frac{2}{s+1}$ with $s\in \N$ defined as the span.
For daily data, letting our span $s=5$ would correspond to a half-life of $5$ days.
The choice of $s$ can be seen as a smoothing factor where high (low) values of $s$ would mean a high (low) degree of smoothing our time series.
Using this method to calculate the well-known exponential weighted moving average of observations $(x_1, \dots, x_n)$ is done by letting $\text{M}^1_{t} = x_t$ for $t = 1, \dots, n$. 
Furthermore, setting $s=2t-1$ would give us the usual average estimate.
Hence, there is a trade-off between the sensitivity to noise and its ability to adapt to parameter changes.


\subsection{Feature Extraction} \label{sec:feature_extraction}


Our interest is to predict risk-adjusted returns, where we incorporate an adjustable hyperparameter that changes the stickiness of the regimes.
We extract the features of our instruments $I$ according to the description of EWMMs in Section \ref{sec:ewmm}.
Denote our features for the first and second moment by $(f^{i}_{s})_{i=1,2} = (\text{EWMM}_{t}^{i})_{i=1,2}$, where $s$ denotes the feature span.
All features $(f^{i}_{s})_{i=1, 2}$ are normalized to zero mean and unit variance using a $z$-score normalization fitted on the training data. 
After normalization, we concatenate our features depending on the moment's order into one feature before passing it onto our HMM.
Thus, our complete features space is $f_{s}=(f^{1}_{s},f^{2}_{s})$.

The span $s$ in our features $f_{s}$ will work as a smoothing factor and determine the frequency of regime shifts, namely the regime stickiness.
The larger we make our smoothing factor $s$, the slower our features $f_{s}$ would change, making our hidden states more sticky, i.e., large diagonal values in the transition matrix $\trans$ (See Section \ref{sec:hmm_elements}).
Thus, portfolio turnover will decrease.

There are different approaches in the literature on how to deal with this increased noise of hidden state prediction; the authors of \cite{gupta2012} use the notion of latency days, in which they forecast the hidden states at time $n+1$ using only the ten previous days of observations.
Others detect a regime change by considering the number of consecutive days in the same new hidden state, given a rolling window of days (which one has to estimate/select). 
Intuitively, smaller window sizes will lead to a larger number of regime changes, whereas large window sizes will increase regimes' length. 
Putting into an economic scenario, one would like to find a window size according to the preferences for turnover adjusted for transaction costs.


\subsection{Prediction of Expected SR} \label{sec:pred_esr}


The unsupervised classification computed by the HMM using our features $f_{s}=(f^{1}_{s},f^{2}_{s})$ results in some mean and variance estimates of every feature in each hidden state $S$.
We aim to combine these resulting mean and variance estimates into a self-explanatory financial metric that reflects the underlying risk-adjusted returns.

Before defining the risk-adjusted return metric we need to introduce the following notions: let $\emissionmu = \{ \mu_j \}_{j=1,\dots ,S}$ denote the mean vectors and $\emissionsigma = \{ \Sigma_j \}_{j=1,\dots ,S}$ the co-variance matrices with $\mu_{j} = (\mu_{j}(f^{1}_{s}), \mu_{j}(f^{2}_{s}))^{T} \in \R^{2}$ and $\Sigma_{j} = \Sigma_{j}(f^{1}_{s},f^{2}_{s}) \in \R^{2 \times 2}$ for $j=1,\dots, S$.
Thus, by dividing our mean estimate of our first moment by the mean estimate of the second moment at each hidden state, we have an Expected SR (ESR) in each hidden state called $\text{ESR}^{j}_{s}$. 
Meaning, for each hidden state $j \in \{ 1,\dots, S \}$, then $\text{ESR}^{j}_{s} = \mu_j(f^{1}_{s})/\mu_j(f^{2}_{s}) \in \R$.
We denote by $\textbf{ESR}^{S}_{s}$ the vector $(\text{ESR}^{1}_{s}, \dots, \text{ESR}^{S}_{s})^T \in \R^{S}$, where $S$ is the number of hidden states in the HMM and $s$ the span used to calculate our features.

We can use our $\textbf{ESR}^{S}_{s}$ metric to predict an expected SR $h \geq 0$ steps ahead by combining this with the estimated vector of state probabilities $\alpha$ and the transition matrix $\trans$.
Recall from \eqref{eq:pred_hs_sequence} that $\alpha_{n+h \vert n} = \alpha_{n \vert n} \trans^{h}$, where $\alpha_{n \vert n}$ is the vector of state probabilities at time $n$ and $\trans$ the transition matrix (given the sequence of observations $(x_1,\dots, x_n)$) with the $j$th entry $(\alpha_{n \vert n})_j = \Prob(z_n=j \vert x)$ for $j=1,\dots, S$.
Hence, we can define the predicted ESR (PESR) metric by the product of
\begin{align} \label{eq:pred_exp_sr}
\text{PESR}^{S}_{s}(h) = (\textbf{ESR}^{S}_{s})^{T} \alpha_{n+h \vert n},
\end{align}
where $h \geq 0$ and $\text{PESR}^{S}_{s}(h) \in \R$. 
This $\text{PESR}^{S}_{s}(h)$ number tells us what SR to expect $h \geq 0$ times ahead.

Summarizing, $\textbf{ESR}^{S}_{s}$ is a vector containing an expected SR of each hidden state of our HMM. 
Thus, by incorporation the transition estimates, we obtain $\text{PESR}^{S}_{s}(h)$ as a metric for predicting expected risk-adjusted returns $h \geq 0$ steps ahead given the HMM with $S$ hidden states. 
Both metrics are fitted on the features using span $s$, extracted from the past observations $(x_1,\dots, x_n)$.
One may note that more elaborating functions could be made by including higher order of moments, incorporating the downside risk of returns. 
Extracting features using closing and opening prices, high and low prices, and volume may also be of interest, as long as the features are not linearly correlated.


\section{Experiments} \label{sec:experiments}


In our experiments, we divide the data set into three parts: training (up to the year 2012 $\approx$ twelve years), validation (the year 2012 to 2016 $\approx$ four years), and test set (from the year 2016 $\approx$ four years).

We train our HMM using the features $f_{s}=(f^{1}_{s},f^{2}_{s})$ extracted from our training data.
Then we validate the (out-of-sample) performance by evaluating our model on the validation data.
Selecting training data with suitable variability will help us improve the models' ability to generalize.
Thus, we identify the desired pattern(s) in our training data, which explains our validation data's behavior the best.
To avoid getting stuck in a local maximum, we select the HMM with the highest score over many trained models, where each model is randomly initialized.

Our goal is to enhance the risk-adjusted returns with the use of our proposed PESR metric $\text{PESR}^{S}_{s}(h)$ from \eqref{eq:pred_exp_sr}.
We choose the number of hidden states relatively low to have high interpretability of each hidden state in our HMM. 
Thus, our choice is an HMM with three hidden states ($S = 3$), where the hidden states can be labeled as a bull, bear, and high volatility regime.
Our labeling comes from the fact that our estimated ESR metric outputs a positive, negative, and (close to) zero value, which can be labeled into a bull, bear, and high volatility regime.
Our high volatility regimes have an estimated ESR metric close to zero as the estimated volatility dominates, i.e., $\mu_{j}(f_{s}^{2})$ is sufficiently larger than $\mu_{j}(f_{s}^{1})$ and $\mu_{j}(f_{s}^{1})$ is close to zero.

We model the outcomes/predictions of the PESR metric $\text{PESR}^{S}_{s}(h)$ into the two different holding strategies; a long-only strategy and long/short strategy.
We will not restrict the turnover level, but we incorporate a transaction cost of 5bps for buying and selling.
Lastly, as we are disallowing gearing, we cap our holdings onto the range $[0,1]$ for the long-only strategy and $[-1,1]$ for the long/short strategy.
If we were to increase the number of hidden states (and/or adding other features) in our HMM, then the PESR metric's outcomes may be transformed into a more advanced holding strategy.

From our training and validation data, we observe that spans $s \in \{15, 30, 60\}$ seems preferable to have some different levels of transitions within the four years of testing. 
Thus, we will in the next section consider span $s \in \{15, 30, 60\}$.
This range of spans $s$ would also illustrate how the choice of span affects our method's regime stickiness.
Recall that the choice of span $s$ will directly affect the turnover, meaning a lower span $s$ may increase (absolute) performance and lower regime stickiness, i.e., increase the level of turnover.

All results in the following section are made using the (out-of-sample) test period from January 2016 to October 2019. 
Before we move to the results of our experiments, then we may need an overview of the instrument's performance metrics to compare with the outcome of our strategies.
In Table \ref{tab:experiments_assets}, we have the annualized returns, annualized volatilities, Sharpe ratios, and maximum drawdowns of each instrument in $I = (I_{1}, \dots, I_{15})^T$.
As we earlier discussed in Section \ref{sec:data}, each instrument's performance metrics vary a lot, but also within each asset class, we have large variations. 
However, most annualized returns are positive (with only a few exceptions) but achieved under different volatility levels.

\begin{table}[!htbp]
\centering
\begin{tabular}{clrrrr}
\hline
\textbf{\#} & \textbf{Instr.} & \textbf{Ret.} & \textbf{Vol.} & \textbf{SR} & \textbf{DD} \\
\hline
1 & CO1 & $ 7.54 \% $ & $ 12.02 \% $ & $ 0.67 $ & $ 7.66 \% $ \\
2 & CO2 & $ -12.94 \% $ & $ 39.25 \% $ & $ -0.16 $ & $ 20.0 \% $ \\
3 & CO3 & $ 10.22 \% $ & $ 33.09 \% $ & $ 0.46 $ & $ 17.44 \% $ \\ \hdashline
4 & FX1 & $ -6.01 \% $ & $ 9.52 \% $ & $ -0.6 $ & $ 9.93 \% $ \\
5 & FX2 & $ -1.68 \% $ & $ 6.84 \% $ & $ -0.21 $ & $ 3.78 \% $ \\
6 & FX3 & $ 1.65 \% $ & $ 8.57 \% $ & $ 0.23 $ & $ 5.56 \% $ \\
7 & FX4 & $ 0.29 \% $ & $ 5.9 \% $ & $ 0.08 $ & $ 3.39 \% $ \\
\hdashline
8 & EQ1 & $ 6.44 \% $ & $ 15.35 \% $ & $ 0.49 $ & $ 11.8 \% $  \\
9 & EQ2 & $ 7.51 \% $ & $ 12.85 \% $ & $ 0.64 $ & $ 6.15 \% $ \\
10 & EQ3 & $ 5.43 \% $ & $ 20.56 \% $ & $ 0.36 $ & $ 14.16 \% $ \\
11 & EQ4 & $ 11.23 \% $ & $ 11.35 \% $ & $ 1.0 $ & $ 7.82 \% $ \\
\hdashline
12 & FI1 & $ 0.92 \% $ & $ 3.78 \% $ & $ 0.27 $ & $ 2.31 \% $ \\
13 & FI2 & $ 4.23 \% $ & $ 4.19 \% $ & $ 1.02 $ & $ 2.1 \% $ \\
14 & FI3 & $ 4.91 \% $ & $ 5.35 \% $ & $ 0.94 $ & $ 3.23 \% $ \\
15 & FI4 & $ 1.59 \% $ & $ 1.71 \% $ & $ 0.96 $ & $ 1.39 \% $ \\
\hline
\end{tabular}
\caption{Realized performance metrics; annualized returns, annualized volatility, Sharpe ratios, and maximum drawdowns of instruments $I$ in the test period from January 2016 to October 2019.}
\label{tab:experiments_assets}
\end{table}


\subsection{Results}


The results of our long-only strategy based on the outcomes of $\text{PESR}^{3}_{s}(1)_{s \in \{ 15, 30, 60 \}}$ are presented in Table \ref{tab:experiments_hmm3_span15_span30_span60_long}.
Table \ref{tab:experiments_hmm3_span15_span30_span60_long} confirms our claim that lower (higher) levels of span $s$ delivers a higher (lower) level of turnover.
However, different choices of span $s$ affect the performance metrics individually due to both the "true" length of market regimes and the transaction costs.
If we consider span $s=30$, then what first comes to mind is that all (annualized) returns are positive with slightly lower (annualized) volatility leading to an improved SR, now above one for all assets (except from FX1, which have a SR of $0.86$).
Furthermore, CO1, EQ4, and FI1, now have a SR above two.
The daily turnover range from $1.64\%$ to $4.08\%$, giving an investment horizon of approximately $25$ to over $60$ days.
Thus, one would have a monthly re-balancing scheme for this long-only strategy.
The overall results presented in Table \ref{tab:experiments_hmm3_span15_span30_span60_long} show a convincing improvement of SR with a feasible turnover rate (which can be changed after preferences through the selection of span $s$).
Nevertheless, we cannot guarantee that the cumulative return will be improved using our PESR metric, as the aim is to improve risk-adjusted returns.
FI4 is an example of this as we see an improved SR but not a cumulative return.
In such cases, additional span sizes should be included to embrace these assets.
Several factors affect the investment strategy, but the choice of span $s$ has a significant influence since it operates as a smoothing factor and determines the regime shifts' frequency (i.e., the regime stickiness). 
Thus, assets with low volatility may not require much smoothing, suggesting that we should use higher levels of span $s$.
In addition, transaction costs play a significant role as the absolute returns are small.

\begin{table*}[!htbp]
\centering
\setlength{\tabcolsep}{2.5pt}
\resizebox{\linewidth}{!}{
\begin{tabular}{cl|rrrrr|rrrrr|rrrrr}
\hline
\multicolumn{2}{c}{\textbf{Long-only}} \vline & \multicolumn{5}{c}{$\textbf{PESR}^{3}_{15}(1)$} \vline & \multicolumn{5}{c}{$\textbf{PESR}^{3}_{30}(1)$} \vline & \multicolumn{5}{c}{$\textbf{PESR}^{3}_{60}(1)$} \\
\hline
\textbf{\#} & \textbf{Instr.} & \textbf{Ret.} & \textbf{Vol.} & \textbf{SR} & \textbf{DD} & \textbf{Turn.} & \textbf{Ret.} & \textbf{Vol.} & \textbf{SR} & \textbf{DD} & \textbf{Turn.} & \textbf{Ret.} & \textbf{Vol.} & \textbf{SR} & \textbf{DD} & \textbf{Turn.} \\
\hline
1 & CO1 
& $ 11.65 \% $ & $ 7.8 \% $ & $ 2.4 $ & $ 6.29 \% $ & $ 4.87 \% $
& $ 13.21 \% $ & $ 8.27 \% $ & $ 2.44 $ & $ 6.48 \% $ & $ 2.67 \% $
& $ 9.61 \% $ & $ 8.39 \% $ & $ 1.74 $ & $ 6.67 \% $ & $ 1.96 \% $ 
\\
2 & CO2 
& $ 29.74 \% $ & $ 24.13 \% $ & $ 2.08 $ & $ 17.7 \% $ & $ 3.87 \% $
& $ 15.56 \% $ & $ 25.29 \% $ & $ 1.2 $ & $ 17.7 \% $ & $ 2.49 \% $ 
& $ 10.57 \% $ & $ 26.39 \% $ & $ 0.73 $ & $ 17.7 \% $ & $ 2.46 \% $
\\
3 & CO3 
& $ 31.57 \% $ & $ 20.62 \% $ & $ 1.95 $ & $ 10.94 \% $ & $ 4.62 \% $
& $ 22.08 \% $ & $ 18.35 \% $ & $ 1.59 $ & $ 12.81 \% $ & $ 3.19 \% $
& $ 15.55 \% $ & $ 17.88 \% $ & $ 1.25 $ & $ 12.81 \% $ & $ 2.54 \% $
\\
\hdashline
4 & FX1
& $ 3.48 \% $ & $ 5.71 \% $ & $ 0.84 $ & $ 4.09 \% $ & $ 4.56 \% $
& $ 2.98 \% $ & $ 4.89 \% $ & $ 0.86 $ & $ 3.67 \% $ & $ 2.52 \% $
& $ 1.65 \% $ & $ 4.51 \% $ & $ 0.55 $ & $ 3.29 \% $ & $ 2.01 \% $
\\
5 & FX2 
& $ 4.63 \% $ & $ 4.26 \% $ & $ 1.93 $ & $ 3.78 \% $ & $ 3.69 \% $
& $ 3.75 \% $ & $ 3.89 \% $ & $ 1.77 $ & $ 2.57 \% $ & $ 2.36 \% $
& $ 2.95 \% $ & $ 3.89 \% $ & $ 1.41 $ & $ 2.57 \% $ & $ 1.47 \% $
\\
6 & FX3 
& $ 4.33 \% $ & $ 5.31 \% $ & $ 1.19 $ & $ 3.59 \% $ & $ 4.02 \% $ 
& $ 4.66 \% $ & $ 5.19 \% $ & $ 1.27 $ & $ 4.72 \% $ & $ 2.7 \% $
& $ 6.73 \% $ & $ 5.75 \% $ & $ 1.54 $ & $ 4.72 \% $ & $ 2.05 \% $
\\
7 & FX4 
& $ 3.37 \% $ & $ 3.16 \% $ & $ 1.94 $ & $ 2.13 \% $ & $ 3.82 \% $ 
& $ 2.89 \% $ & $ 3.7 \% $ & $ 1.29 $ & $ 2.71 \% $ & $ 2.46 \% $
& $ 3.44 \% $ & $ 4.13 \% $ & $ 1.08 $ & $ 2.42 \% $ & $ 2.06 \% $
\\
\hdashline
8 & EQ1
& $ 15.64 \% $ & $ 8.56 \% $ & $ 2.48 $ & $ 11.32 \% $ & $ 4.49 \% $
& $ 11.51 \% $ & $ 8.92 \% $ & $ 1.73 $ & $ 11.32 \% $ & $ 3.74 \% $
& $ 8.04 \% $ & $ 9.84 \% $ & $ 1.04 $ & $ 11.32 \% $ & $ 2.66 \% $
\\
9 & EQ2 
& $ 13.44 \% $ & $ 6.97 \% $ & $ 2.55 $ & $ 4.63 \% $ & $ 4.91 \% $
& $ 8.02 \% $ & $ 7.68 \% $ & $ 1.36 $ & $ 4.91 \% $ & $ 3.77 \% $
& $ 9.81 \% $ & $ 8.18 \% $ & $ 1.54 $ & $ 5.67 \% $ & $ 2.24 \% $
\\
10 & EQ3 
& $ 23.22 \% $ & $ 11.22 \% $ & $ 2.46 $ & $ 7.58 \% $ & $ 5.4 \% $
& $ 16.85 \% $ & $ 11.47 \% $ & $ 1.8 $ & $ 7.58 \% $ & $ 3.71 \% $
& $ 15.42 \% $ & $ 12.05 \% $ & $ 1.58 $ & $ 9.58 \% $ & $ 2.06 \% $
\\
11 & EQ4 
& $ 14.4 \% $ & $ 6.35 \% $ & $ 3.0 $ & $ 5.17 \% $ & $ 4.79 \% $
& $ 12.59 \% $ & $ 7.17 \% $ & $ 2.01 $ & $ 7.72 \% $ & $ 3.43 \% $ 
& $ 10.93 \% $ & $ 8.28 \% $ & $ 1.44 $ & $ 7.72 \% $ & $ 2.07 \% $
\\
\hdashline
12 & FI1
& $ 2.01 \% $ & $ 1.91 \% $ & $ 2.49 $ & $ 1.65 \% $ & $ 2.44 \% $
& $ 2.26 \% $ & $ 1.97 \% $ & $ 2.58 $ & $ 1.69 \% $ & $ 1.64 \% $
& $ 1.83 \% $ & $ 1.99 \% $ & $ 1.92 $ & $ 1.82 \% $ & $ 1.44 \% $
\\
13 & FI2 
& $ 5.21 \% $ & $ 2.74 \% $ & $ 2.62 $ & $ 1.53 \% $ & $ 4.91 \% $
& $ 4.09 \% $ & $ 2.85 \% $ & $ 1.93 $ & $ 2.07 \% $ & $ 4.08 \% $
& $ 3.72 \% $ & $ 3.01 \% $ & $ 1.6 $ & $ 2.07 \% $ & $ 2.97 \% $
\\
14 & FI3 
& $ 6.0 \% $ & $ 3.97 \% $ & $ 1.8 $ & $ 2.18 \% $ & $ 4.64 \% $
& $ 4.17 \% $ & $ 3.71 \% $ & $ 1.35 $ & $ 2.22 \% $ & $ 3.87 \% $
& $ 3.86 \% $ & $ 3.73 \% $ & $ 1.31 $ & $ 3.1 \% $ & $ 3.09 \% $
\\
15 & FI4 
& $ 0.5 \% $ & $ 1.04 \% $ & $ 1.25 $ & $ 1.39 \% $ & $ 1.63 \% $
& $ 0.74 \% $ & $ 1.01 \% $ & $ 1.62 $ & $ 1.12 \% $ & $ 1.68 \% $
& $ 1.36 \% $ & $ 1.21 \% $ & $ 1.97 $ & $ 1.39 \% $ & $ 1.27 \% $
\\
\hline
\end{tabular}
}
\caption{Realized performance metrics; annualized returns, annualized volatilities, Sharpe ratios, maximum drawdowns, and daily turnovers of long-only strategies $\text{PESR}^{3}_{s}(1)_{s \in \{ 15, 30, 60 \}}$ in the test period from January 2016 to October 2019.}
\label{tab:experiments_hmm3_span15_span30_span60_long}
\end{table*}

Next, in Table \ref{tab:experiments_hmm3_span15_span30_span60_longshort}, we have the results of our long/short strategy; this strategy seems to provide larger (absolute) returns but with increased volatility, leading to a lower SR than for the long-only strategy.
This means the short leg of our strategies adds some more volatility to the strategy.
Naturally, as we can be short now, this leads to increasing daily turnover, e.g., for span $s=30$, the turnover now ranges from $3.37\%$ to $7.34\%$ giving an investment horizon of approximately $15$ to $30$ days.
As the turnover increase, the same do transaction costs, which for some strategies/assets may represent a significant part of the overall performance.
Particularly, FI4 has a negative SR (and cumulative return), however, with lower volatility than the asset itself.

\begin{table*}[!htbp]
\centering
\setlength{\tabcolsep}{2.5pt}
\resizebox{\linewidth}{!}{
\begin{tabular}{cl|rrrrr|rrrrr|rrrrr}
\hline
\multicolumn{2}{c}{\textbf{Long/short}} \vline & \multicolumn{5}{c}{$\textbf{PESR}^{3}_{15}(1)$} \vline & \multicolumn{5}{c}{$\textbf{PESR}^{3}_{30}(1)$} \vline & \multicolumn{5}{c}{$\textbf{PESR}^{3}_{60}(1)$} \\
\hline
\textbf{\#} & \textbf{Instr.} & \textbf{Ret.} & \textbf{Vol.} & \textbf{SR} & \textbf{DD} & \textbf{Turn.} & \textbf{Ret.} & \textbf{Vol.} & \textbf{SR} & \textbf{DD} & \textbf{Turn.} & \textbf{Ret.} & \textbf{Vol.} & \textbf{SR} & \textbf{DD} & \textbf{Turn.} \\
\hline
1 & CO1
& $ 15.69 \% $ & $ 11.53 \% $ & $ 1.31 $ & $ 7.76 \% $ & $ 9.68 \% $ 
& $ 18.45 \% $ & $ 11.62 \% $ & $ 1.52 $ & $ 6.66 \% $ & $ 5.22 \% $
& $ 9.94 \% $ & $ 9.27 \% $ & $ 1.07 $ & $ 6.67 \% $ & $ 3.13 \% $
\\
2 & CO2
& $ 51.61 \% $ & $ 33.75 \% $ & $ 1.4 $ & $ 17.7 \% $ & $ 7.61 \% $ 
& $ 33.52 \% $ & $ 34.48 \% $ & $ 1.01 $ & $ 17.7 \% $ & $ 4.72 \% $
& $ 26.5 \% $ & $ 36.1 \% $ & $ 0.83 $ & $ 17.7 \% $ & $ 4.81 \% $
\\
3 & CO3 
& $ 53.3 \% $ & $ 27.25 \% $ & $ 1.7 $ & $ 11.4 \% $ & $ 9.03 \% $
& $ 26.26 \% $ & $ 26.62 \% $ & $ 1.01 $ & $ 13.88 \% $ & $ 6.63 \% $
& $ 16.58 \% $ & $ 26.12 \% $ & $ 0.72 $ & $ 13.11 \% $ & $ 5.27 \% $
\\
\hdashline
4 & FX1 
& $ 10.74 \% $ & $ 7.25 \% $ & $ 1.43 $ & $ 4.09 \% $ & $ 8.22 \% $ 
& $ 9.38 \% $ & $ 6.96 \% $ & $ 1.32 $ & $ 3.67 \% $ & $ 5.11 \% $
& $ 6.77 \% $ & $ 7.13 \% $ & $ 0.95 $ & $ 3.93 \% $ & $ 3.9 \% $
\\
5 & FX2 
& $ 10.64 \% $ & $ 6.64 \% $ & $ 1.55 $ & $ 3.78 \% $ & $ 7.56 \% $ 
& $ 9.62 \% $ & $ 6.52 \% $ & $ 1.44 $ & $ 3.31 \% $ & $ 4.74 \% $
& $ 8.03 \% $ & $ 6.54 \% $ & $ 1.22 $ & $ 3.31 \% $ & $ 2.98 \% $
\\
6 & FX3 
& $ 7.42 \% $ & $ 6.89 \% $ & $ 1.06 $ & $ 3.98 \% $ & $ 7.38 \% $ 
& $ 7.68 \% $ & $ 6.64 \% $ & $ 1.15 $ & $ 5.02 \% $ & $ 4.92 \% $
& $ 9.51 \% $ & $ 6.9 \% $ & $ 1.35 $ & $ 5.02 \% $ & $ 4.1 \% $
\\
7 & FX4
& $ 6.05 \% $ & $ 5.68 \% $ & $ 1.06 $ & $ 3.74 \% $ & $ 7.72 \% $ 
& $ 5.34 \% $ & $ 5.7 \% $ & $ 0.94 $ & $ 3.32 \% $ & $ 5.0 \% $
& $ 5.91 \% $ & $ 5.71 \% $ & $ 1.04 $ & $ 3.32 \% $ & $ 4.16 \% $
\\
\hdashline
8 & EQ1
& $ 27.3 \% $ & $ 13.5 \% $ & $ 1.87 $ & $ 11.68 \% $ & $ 8.66 \% $ 
& $ 13.81 \% $ & $ 13.16 \% $ & $ 1.06 $ & $ 11.32 \% $ & $ 7.18 \% $
& $ 10.74 \% $ & $ 13.88 \% $ & $ 0.81 $ & $ 12.07 \% $ & $ 5.02 \% $
\\
9 & EQ2 
& $ 23.33 \% $ & $ 11.27 \% $ & $ 1.93 $ & $ 6.72 \% $ & $ 9.85 \% $
& $ 13.14 \% $ & $ 11.36 \% $ & $ 1.15 $ & $ 6.6 \% $ & $ 7.34 \% $
& $ 11.61 \% $ & $ 11.99 \% $ & $ 0.98 $ & $ 6.72 \% $ & $ 4.58 \% $
\\
10 & EQ3
& $ 43.07 \% $ & $ 16.19 \% $ & $ 2.3 $ & $ 8.15 \% $ & $ 10.64 \% $
& $ 29.11 \% $ & $ 16.81 \% $ & $ 1.62 $ & $ 12.54 \% $ & $ 6.98 \% $
& $ 22.73 \% $ & $ 16.86 \% $ & $ 1.31 $ & $ 12.15 \% $ & $ 4.43 \% $
\\
11 & EQ4 
& $ 18.51 \% $ & $ 10.55 \% $ & $ 1.66 $ & $ 6.65 \% $ & $ 9.16 \% $
& $ 15.01 \% $ & $ 10.66 \% $ & $ 1.36 $ & $ 9.17 \% $ & $ 6.51 \% $
& $ 10.67 \% $ & $ 11.06 \% $ & $ 0.97 $ & $ 7.86 \% $ & $ 3.96 \% $
\\
\hdashline
12 & FI1
& $ 3.28 \% $ & $ 3.68 \% $ & $ 0.9 $ & $ 1.81 \% $ & $ 4.74 \% $ 
& $ 3.58 \% $ & $ 3.67 \% $ & $ 0.99 $ & $ 1.87 \% $ & $ 3.37 \% $
& $ 2.32 \% $ & $ 2.66 \% $ & $ 0.87 $ & $ 1.82 \% $ & $ 2.46 \% $
\\
13 & FI2
& $ 6.28 \% $ & $ 4.0 \% $ & $ 1.52 $ & $ 1.98 \% $ & $ 9.92 \% $ 
& $ 4.93 \% $ & $ 4.0 \% $ & $ 1.21 $ & $ 2.07 \% $ & $ 7.48 \% $
& $ 3.35 \% $ & $ 3.59 \% $ & $ 0.93 $ & $ 2.07 \% $ & $ 5.07 \% $
\\
14 & FI3
& $ 6.98 \% $ & $ 4.66 \% $ & $ 1.47 $ & $ 2.33 \% $ & $ 9.13 \% $ 
& $ 4.24 \% $ & $ 4.41 \% $ & $ 0.97 $ & $ 2.36 \% $ & $ 6.92 \% $
& $ 3.49 \% $ & $ 4.45 \% $ & $ 0.8 $ & $ 3.15 \% $ & $ 5.31 \% $
\\
15 & FI4
& $ -0.57 \% $ & $ 1.66 \% $ & $ -0.34 $ & $ 1.39 \% $ & $ 3.29 \% $ 
& $ -0.2 \% $ & $ 1.57 \% $ & $ -0.12 $ & $ 1.12 \% $ & $ 3.45 \% $
& $ 1.22 \% $ & $ 1.35 \% $ & $ 0.91 $ & $ 1.39 \% $ & $ 2.03 \% $
\\
\hline
\end{tabular}
}
\caption{Realized performance metrics; annualized returns, annualized volatilities, Sharpe ratios, maximum drawdowns, and daily turnovers of long/short strategies $\text{PESR}^{3}_{s}(1)_{s \in \{ 15, 30, 60 \}}$ in the test period from January 2016 to October 2019.}
\label{tab:experiments_hmm3_span15_span30_span60_longshort}
\end{table*}

Time-series plots of cumulative returns of each instrument $I$ for both HMM strategies (long-only and long/short) can be found in \ref{sec:appendix_strat}, including their corresponding holdings.
These figures show that we mostly shift between the bull and bear regime, and only in the high volatility state for some short periods.
Overall, as we seek to increase our risk-adjusted returns, then the long-only strategy would be preferred.
However, if we relaxed our risk-aversion, we could maximize total return using the long/short strategy.


\section{Discussion} \label{sec:conclusion}


HMMs have previously been applied to finance time series with great success but never on a broad class of assets, at least not to our knowledge.
We proposed an asset independent three-state HMM for predicting risk-adjusted returns trained using only the first two moments as features.
The model outcome was combined into a metric for predicting expected SRs.
Our investigation showed a proper ability to predict bull, bear, and high-volatility regimes, which lead to enhanced risk-adjusted returns (compared to buying the underlying asset) while keeping a preferable turnover level.
However, this could be improved by fine-tuning the choice of span $s$ as transaction costs could otherwise dominate.

As our findings were made using the entire test dataset to predict the hidden state sequence, our next focus will then be an extension to a setting in which we make incremental predictions of tomorrow's expected SR using only past information.
As this may increase noise, we could increase our model's predictability by introducing time-varying parameters, i.e., an adaptive model where the model parameters are updated as new observations arrive (e.g., see \cite{ford1998} and \cite{nystrup2017}).
Expanding this analysis with a larger group of features, e.g., volume, higher-order moments, short-term oscillators, and associated gradients, could be appealing.
All this could be combined with the feature saliency HMM proposed by \cite{adams2016}, which comprises the treatment of "irrelevant" features.


\section*{Acknowledgement}


This work was supported by a grant from R\'egion Ile de France.
I owe Advestis a great appreciation for their devotion and willingness to spend generous amounts of time with me.
I also owe them gratitude for providing the anonymized data used in the analysis.
I would very much like to acknowledge Christophe Geissler, Vincent Margot, and Nicolas Morizet for our constructive discussions and their valuable suggestions during this research.


\bibliographystyle{elsarticle-harv} 
\bibliography{bibliography}



\appendix



\section{Cumulative Returns of HMM Strategies} \label{sec:appendix_strat}

\begin{figure}[!htbp]
\centering
\includegraphics[width=\figscale\linewidth]{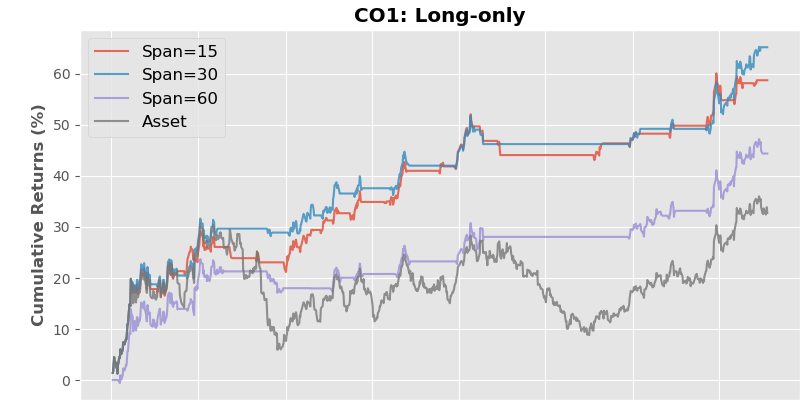} 
\includegraphics[width=\figscale\linewidth]{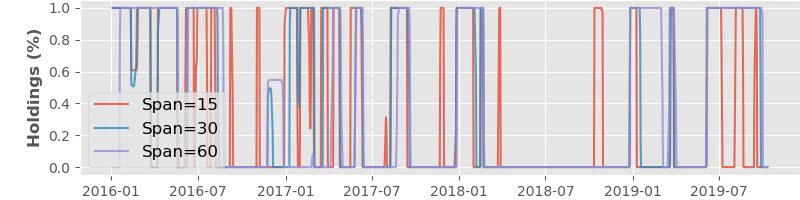}
\caption{Cumulative returns of long-only HMM strategies for instrument $I_{1}$ (CO1) in the test period from January 2016 to October 2019.}
\label{fig:strat_l_gc1}
\end{figure}
\begin{figure}[!htbp]
\centering
\includegraphics[width=\figscale\linewidth]{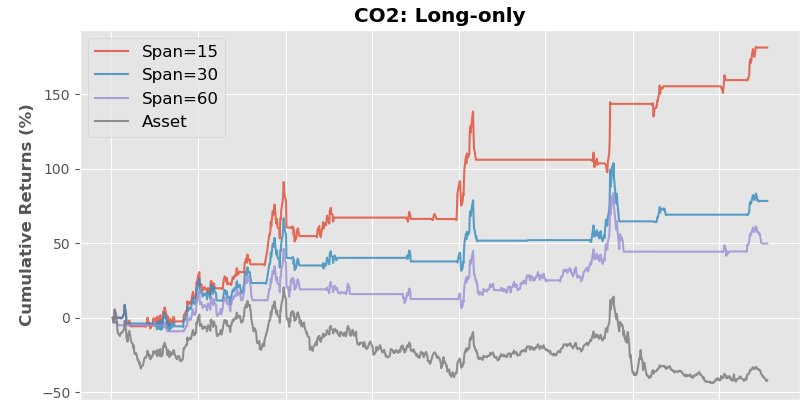} 
\includegraphics[width=\figscale\linewidth]{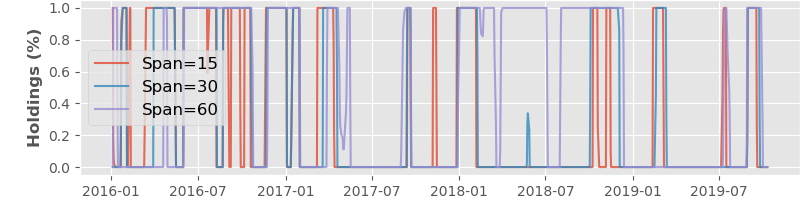}
\caption{Cumulative returns of long-only HMM strategies for instrument $I_{2}$ (CO2) in the test period from January 2016 to October 2019.}
\label{fig:strat_l_ng1}
\end{figure}
\begin{figure}[!htbp]
\centering
\includegraphics[width=\figscale\linewidth]{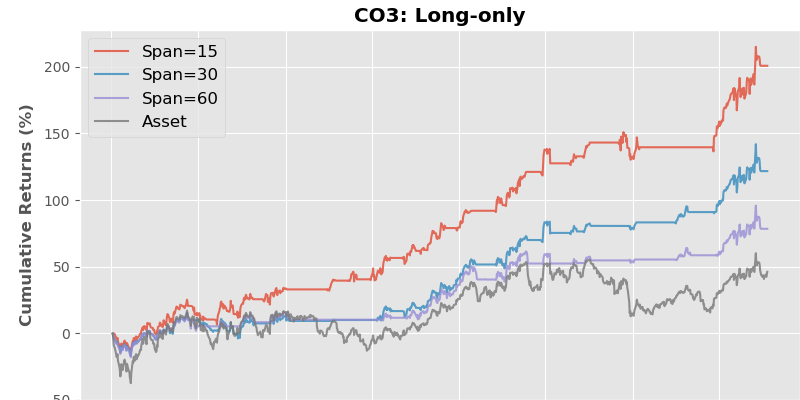} 
\includegraphics[width=\figscale\linewidth]{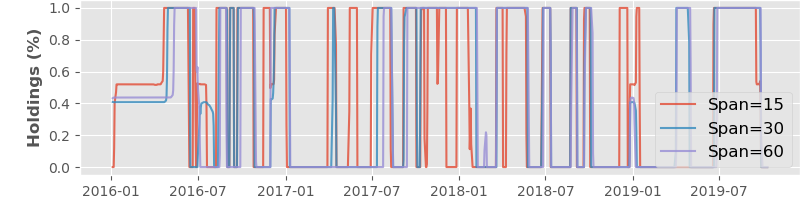}
\caption{Cumulative returns of long-only HMM strategies for instrument $I_{3}$ (CO3) in the test period from January 2016 to October 2019.}
\label{fig:strat_l_cl1}
\end{figure}
\begin{figure}[!htbp]
\centering
\includegraphics[width=\figscale\linewidth]{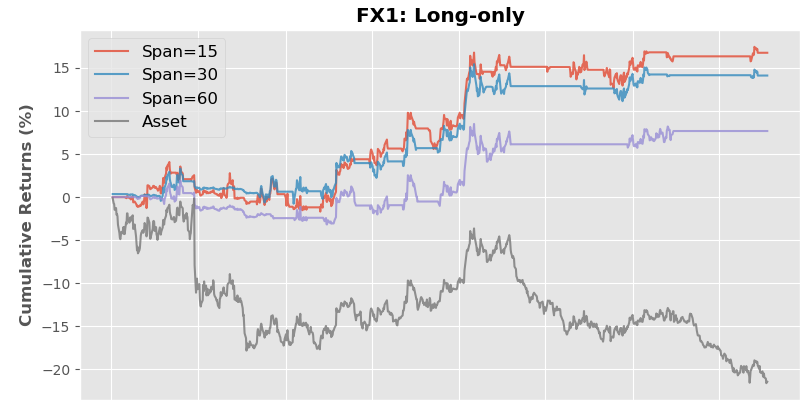} 
\includegraphics[width=\figscale\linewidth]{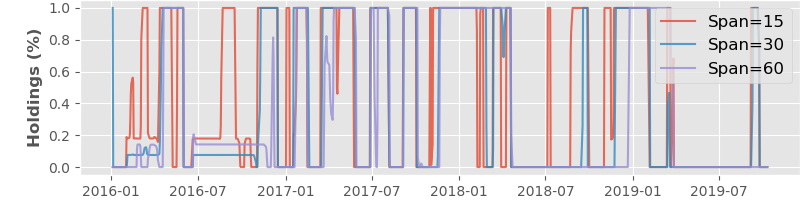}
\caption{Cumulative returns of long-only HMM strategies for instrument $I_{4}$ (FX1) in the test period from January 2016 to October 2019.}
\label{fig:strat_l_bp1}
\end{figure}
\begin{figure}[!htbp]
\centering
\includegraphics[width=\figscale\linewidth]{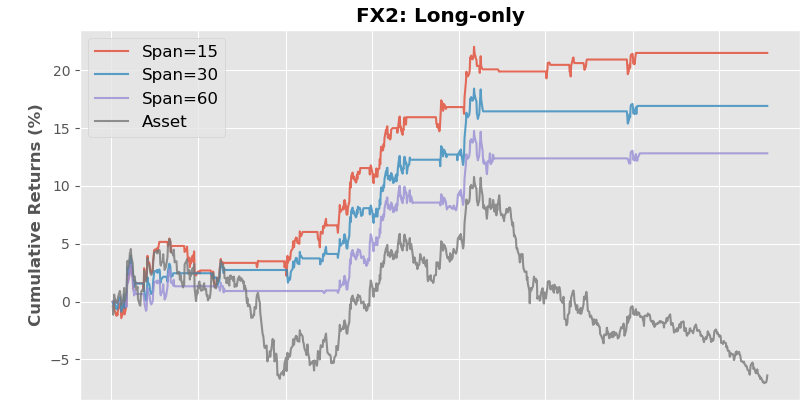} 
\includegraphics[width=\figscale\linewidth]{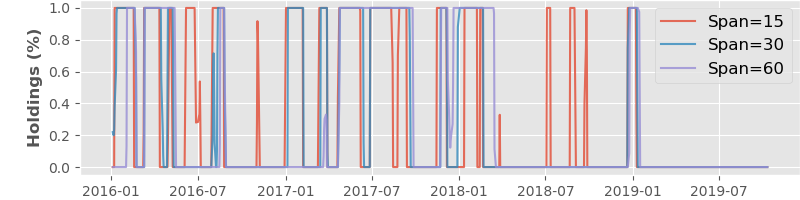}
\caption{Cumulative returns of long-only HMM strategies for instrument $I_{5}$ (FX2) in the test period from January 2016 to October 2019.}
\label{fig:strat_l_ec1}
\end{figure}
\begin{figure}[!htbp]
\centering
\includegraphics[width=\figscale\linewidth]{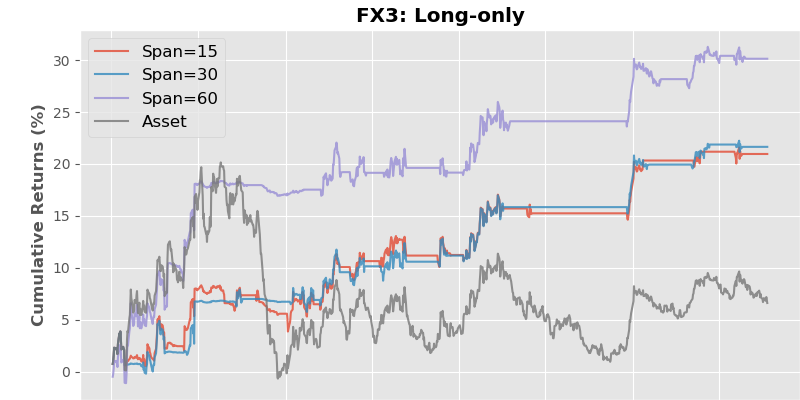} 
\includegraphics[width=\figscale\linewidth]{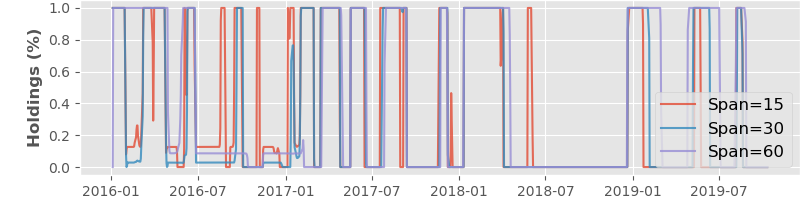}
\caption{Cumulative returns of long-only HMM strategies for instrument $I_{6}$ (FX3) in the test period from January 2016 to October 2019.}
\label{fig:strat_l_jy1}
\end{figure}
\begin{figure}[!htbp]
\centering
\includegraphics[width=\figscale\linewidth]{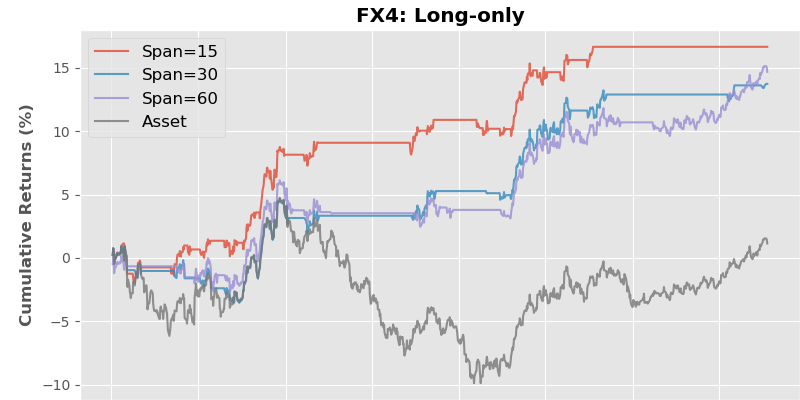} 
\includegraphics[width=\figscale\linewidth]{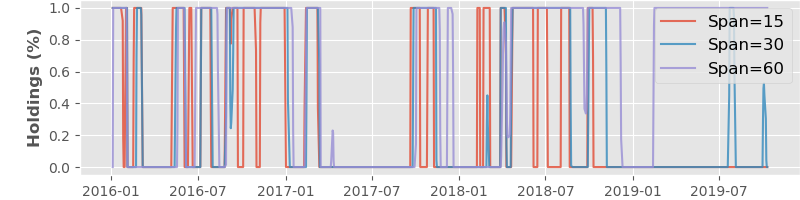}
\caption{Cumulative returns of long-only HMM strategies for instrument $I_{7}$ (FX4) in the test period from January 2016 to October 2019.}
\label{fig:strat_l_dx1}
\end{figure}
\begin{figure}[!htbp]
\centering
\includegraphics[width=\figscale\linewidth]{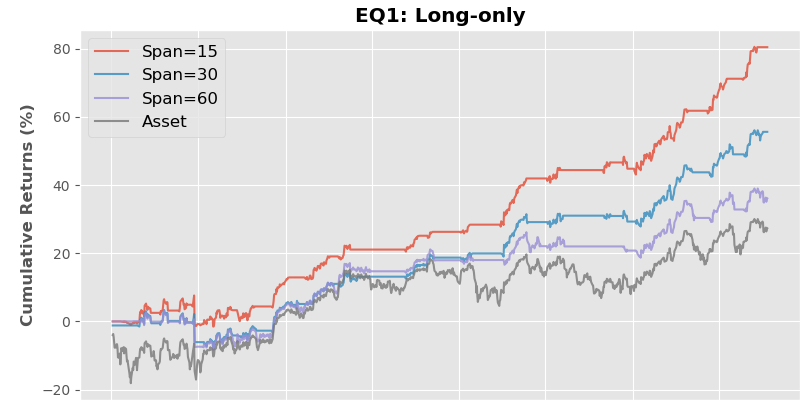} 
\includegraphics[width=\figscale\linewidth]{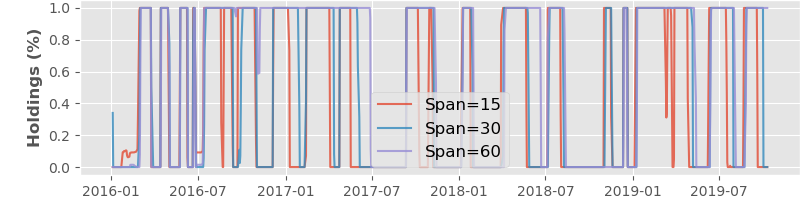}
\caption{Cumulative returns of long-only HMM strategies for instrument $I_{8}$ (EQ1) in the test period from January 2016 to October 2019.}
\label{fig:strat_l_vg1}
\end{figure}
\begin{figure}[!htbp]
\centering
\includegraphics[width=\figscale\linewidth]{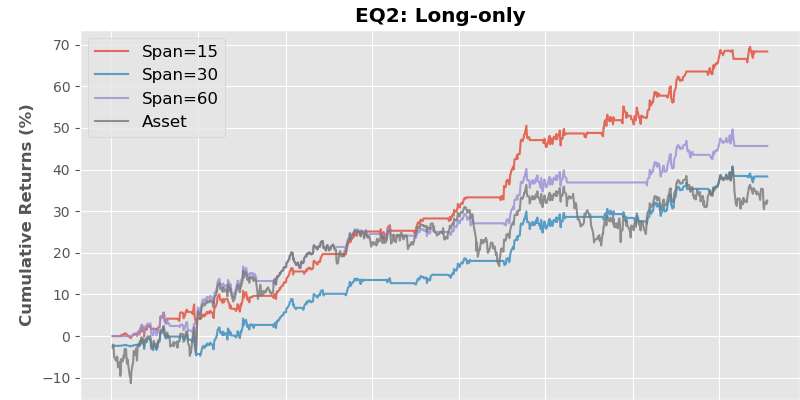} 
\includegraphics[width=\figscale\linewidth]{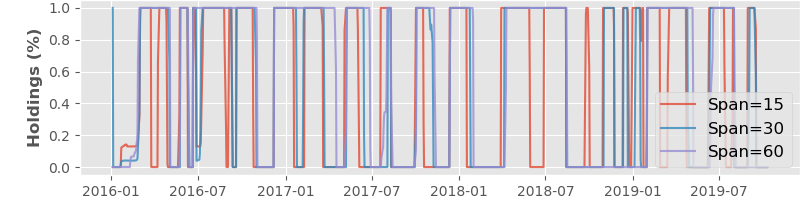}
\caption{Cumulative returns of long-only HMM strategies for instrument $I_{9}$ (EQ2) in the test period from January 2016 to October 2019.}
\label{fig:strat_l_z_1}
\end{figure}
\begin{figure}[!htbp]
\centering
\includegraphics[width=\figscale\linewidth]{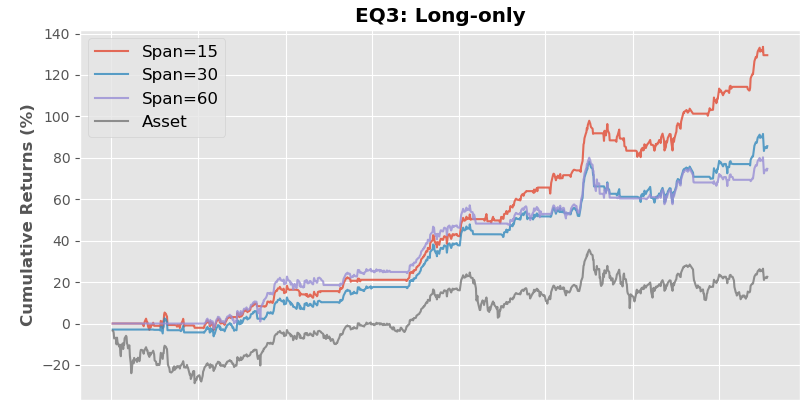} 
\includegraphics[width=\figscale\linewidth]{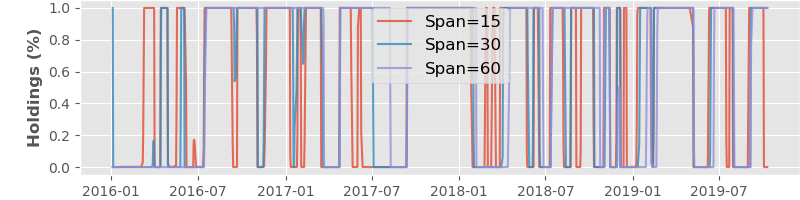}
\caption{Cumulative returns of long-only HMM strategies for instrument $I_{10}$ (EQ3) in the test period from January 2016 to October 2019.}
\label{fig:strat_l_nk1}
\end{figure}
\begin{figure}[!htbp]
\centering
\includegraphics[width=\figscale\linewidth]{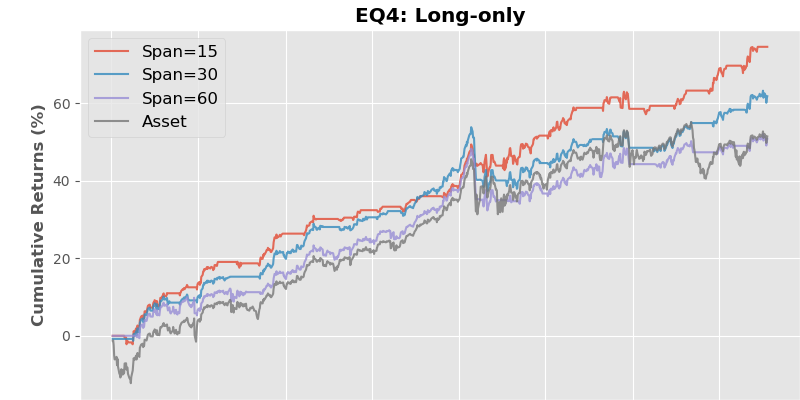} 
\includegraphics[width=\figscale\linewidth]{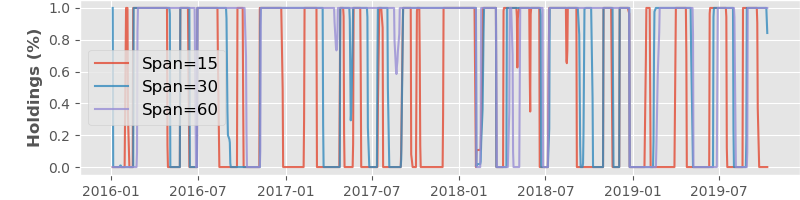}
\caption{Cumulative returns of long-only HMM strategies for instrument $I_{11}$ (EQ4) in the test period from January 2016 to October 2019.}
\label{fig:strat_l_es1}
\end{figure}
\begin{figure}[!htbp]
\centering
\includegraphics[width=\figscale\linewidth]{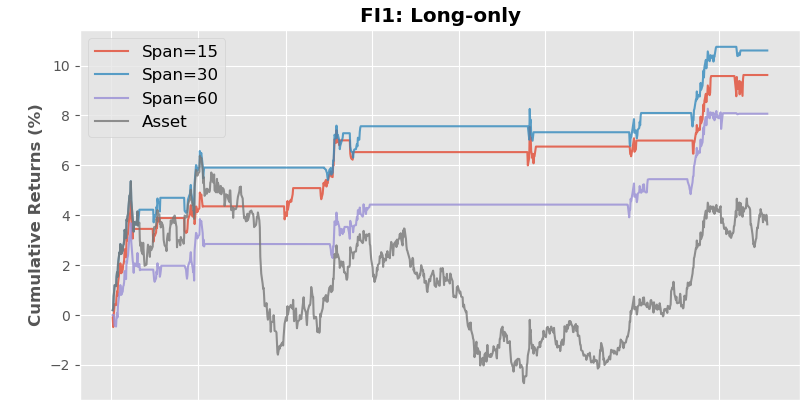} 
\includegraphics[width=\figscale\linewidth]{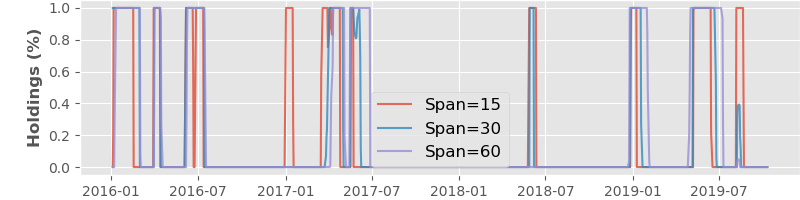}
\caption{Cumulative returns of long-only HMM strategies for instrument $I_{12}$ (FI1) in the test period from January 2016 to October 2019.}
\label{fig:strat_l_ty1}
\end{figure}
\begin{figure}[!htbp]
\centering
\includegraphics[width=\figscale\linewidth]{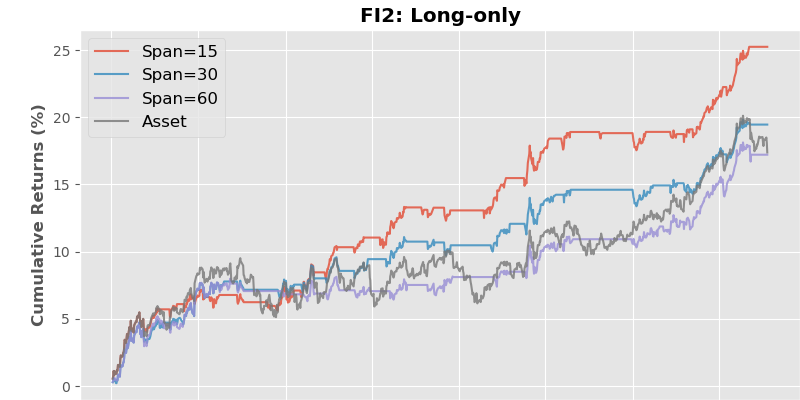} 
\includegraphics[width=\figscale\linewidth]{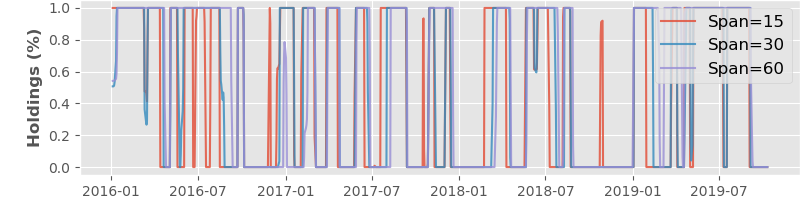}
\caption{Cumulative returns of long-only HMM strategies for instrument $I_{13}$ (FI2) in the test period from January 2016 to October 2019.}
\label{fig:strat_l_rx1}
\end{figure}
\begin{figure}[!htbp]
\centering
\includegraphics[width=\figscale\linewidth]{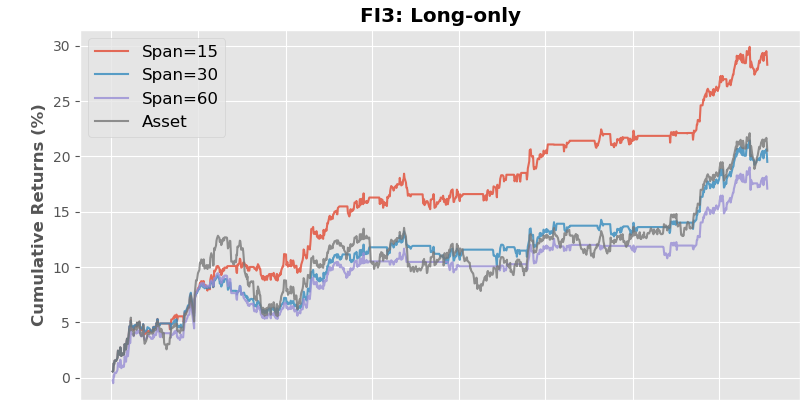} 
\includegraphics[width=\figscale\linewidth]{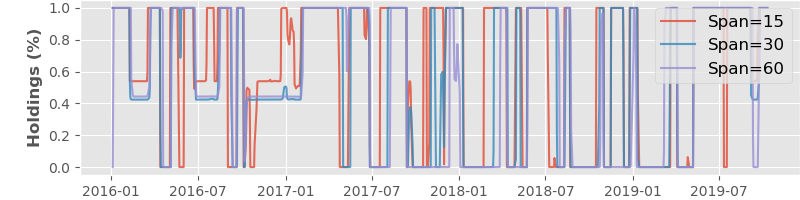}
\caption{Cumulative returns of long-only HMM strategies for instrument $I_{14}$ (FI3) in the test period from January 2016 to October 2019.}
\label{fig:strat_l_g_1}
\end{figure}
\begin{figure}[!htbp]
\centering
\includegraphics[width=\figscale\linewidth]{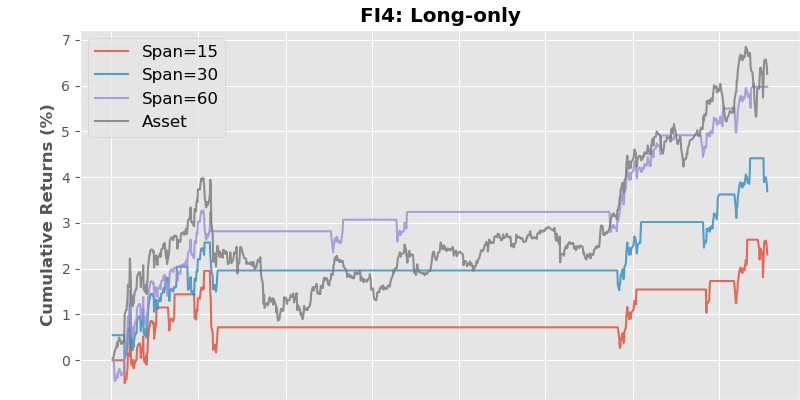} 
\includegraphics[width=\figscale\linewidth]{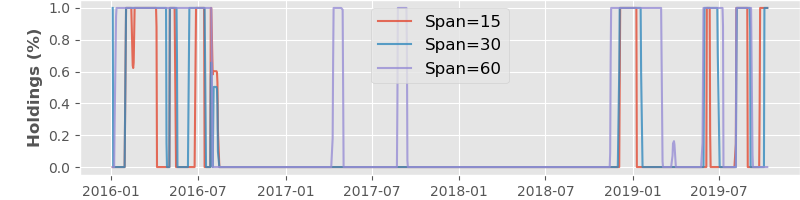}
\caption{Cumulative returns of long-only HMM strategies for instrument $I_{15}$ (FI4) in the test period from January 2016 to October 2019.}
\label{fig:strat_l_jb1}
\end{figure}
\begin{figure}[!htbp]
\centering
\includegraphics[width=\figscale\linewidth]{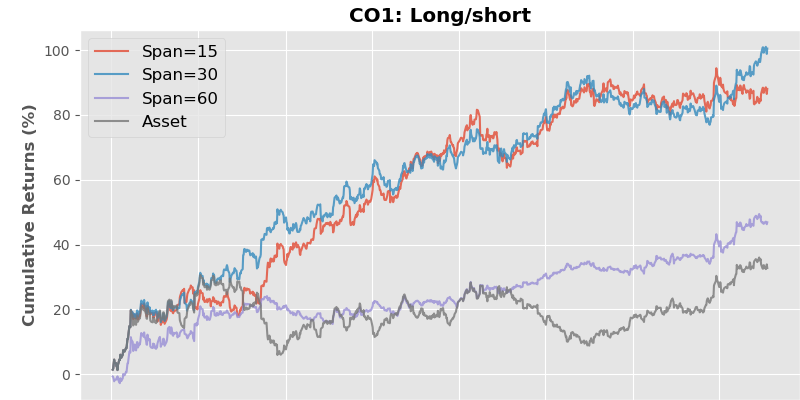} 
\includegraphics[width=\figscale\linewidth]{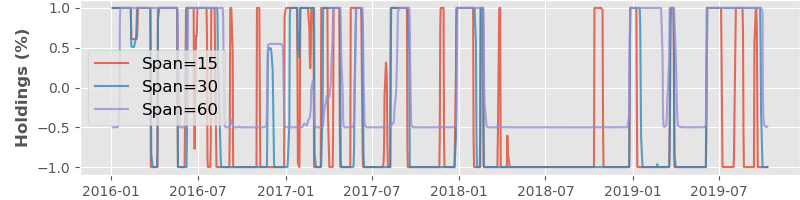}
\caption{Cumulative returns of long/short HMM strategies for instrument $I_{1}$ (CO1) in the test period from January 2016 to October 2019.}
\label{fig:strat_ls_gc1}
\end{figure}
\begin{figure}[!htbp]
\centering
\includegraphics[width=\figscale\linewidth]{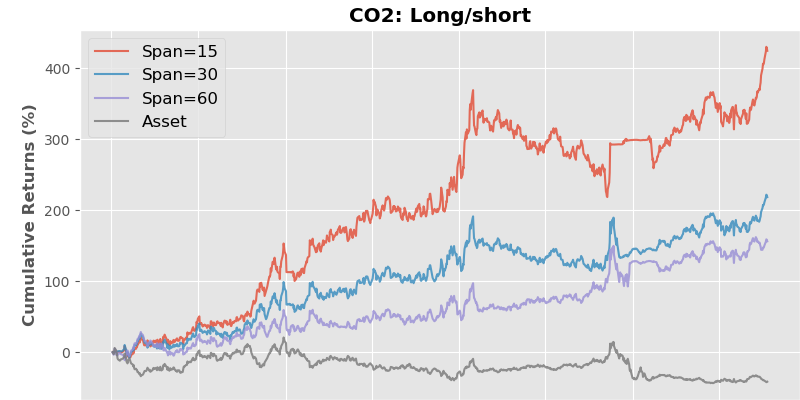} 
\includegraphics[width=\figscale\linewidth]{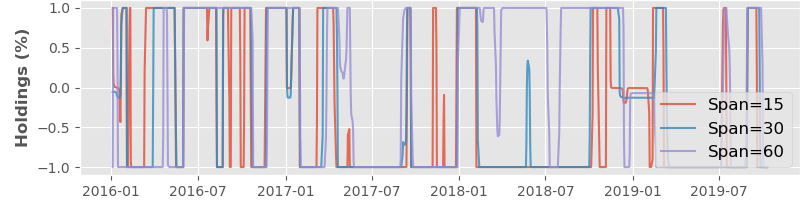}
\caption{Cumulative returns of long/short HMM strategies for instrument $I_{2}$ (CO2) in the test period from January 2016 to October 2019.}
\label{fig:strat_ls_ng1}
\end{figure}
\begin{figure}[!htbp]
\centering
\includegraphics[width=\figscale\linewidth]{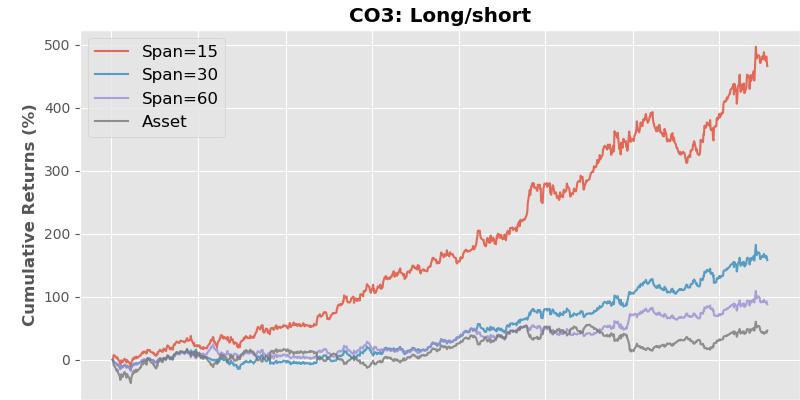} 
\includegraphics[width=\figscale\linewidth]{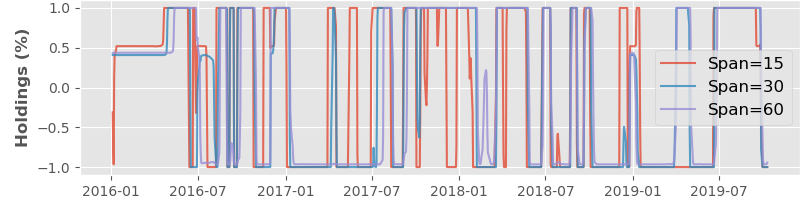}
\caption{Cumulative returns of long/short HMM strategies for instrument $I_{3}$ (CO3) in the test period from January 2016 to October 2019.}
\label{fig:strat_ls_cl1}
\end{figure}
\begin{figure}[!htbp]
\centering
\includegraphics[width=\figscale\linewidth]{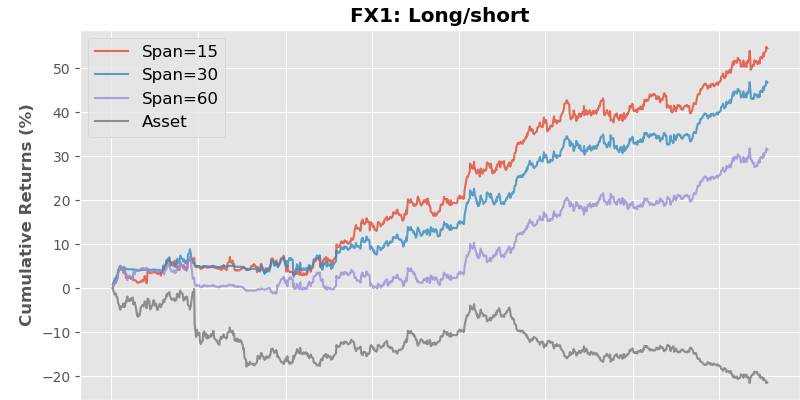} 
\includegraphics[width=\figscale\linewidth]{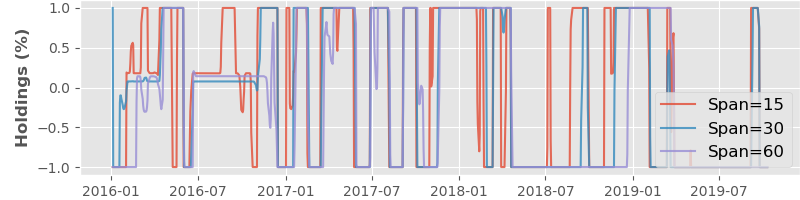}
\caption{Cumulative returns of long/short HMM strategies for instrument $I_{4}$ (FX1) in the test period from January 2016 to October 2019.}
\label{fig:strat_ls_bp1}
\end{figure}
\begin{figure}[!htbp]
\centering
\includegraphics[width=\figscale\linewidth]{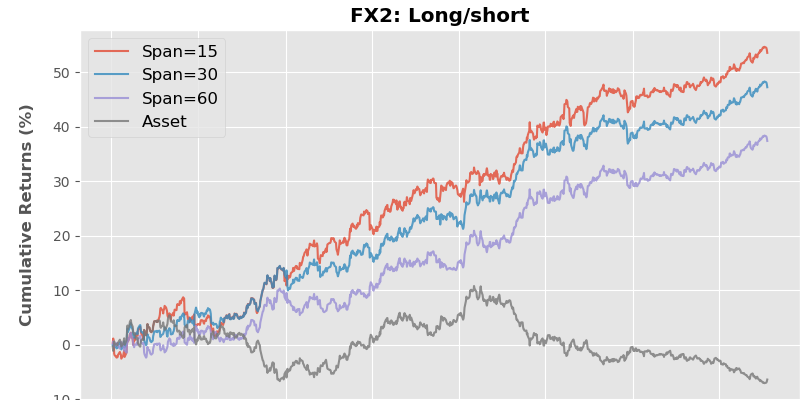} 
\includegraphics[width=\figscale\linewidth]{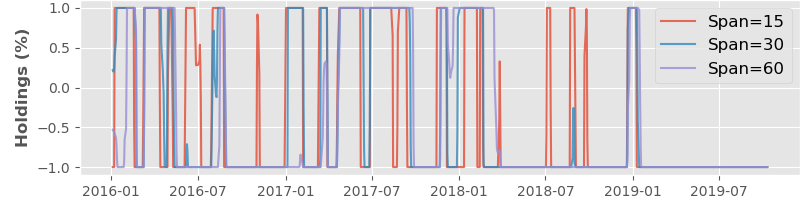}
\caption{Cumulative returns of long/short HMM strategies for instrument $I_{5}$ (FX2) in the test period from January 2016 to October 2019.}
\label{fig:strat_ls_ec1}
\end{figure}
\begin{figure}[!htbp]
\centering
\includegraphics[width=\figscale\linewidth]{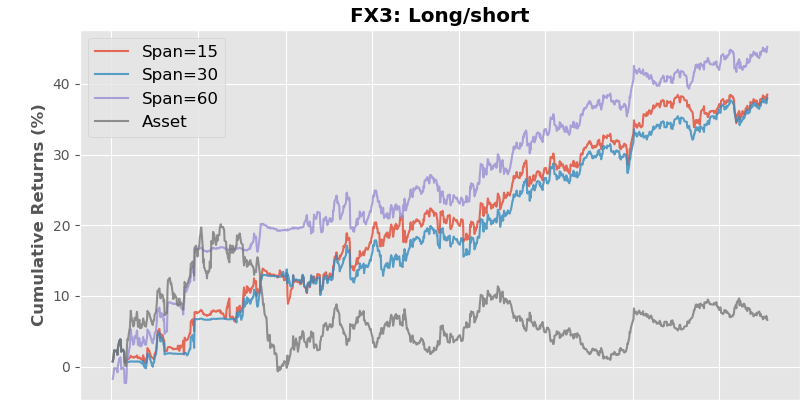} 
\includegraphics[width=\figscale\linewidth]{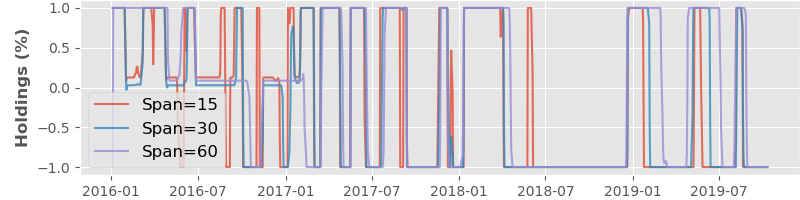}
\caption{Cumulative returns of long/short HMM strategies for instrument $I_{6}$ (FX3) in the test period from January 2016 to October 2019.}
\label{fig:strat_ls_jy1}
\end{figure}
\begin{figure}[!htbp]
\centering
\includegraphics[width=\figscale\linewidth]{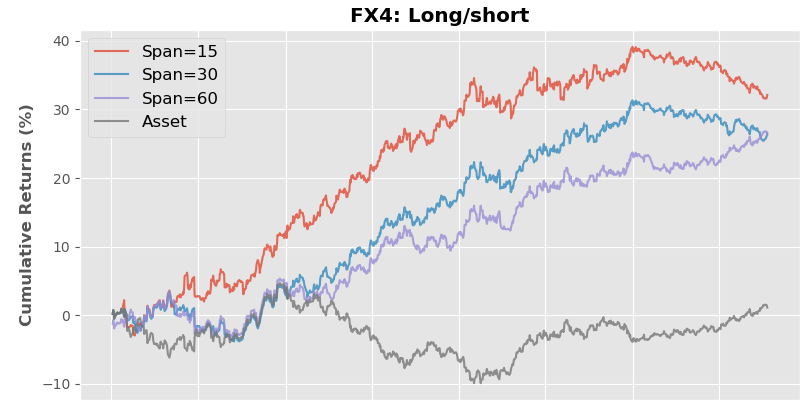} 
\includegraphics[width=\figscale\linewidth]{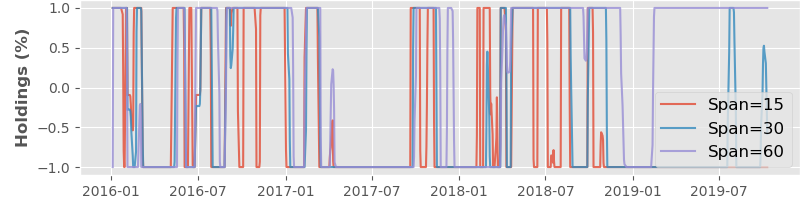}
\caption{Cumulative returns of long/short HMM strategies for instrument $I_{7}$ (FX4) in the test period from January 2016 to October 2019.}
\label{fig:strat_ls_dx1}
\end{figure}
\begin{figure}[!htbp]
\centering
\includegraphics[width=\figscale\linewidth]{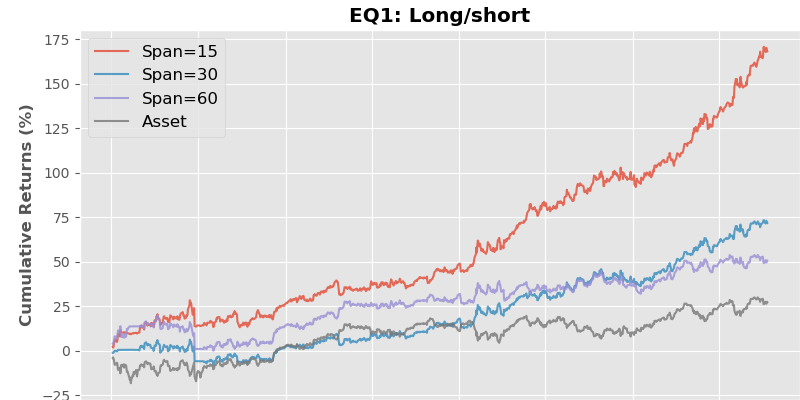} 
\includegraphics[width=\figscale\linewidth]{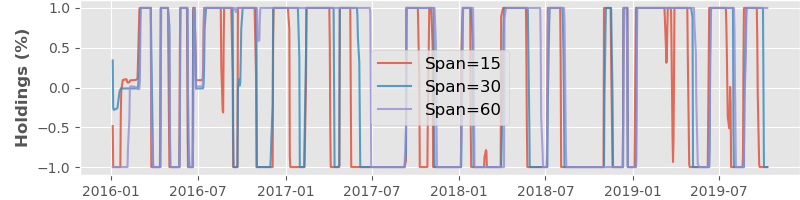}
\caption{Cumulative returns of long/short HMM strategies for instrument $I_{8}$ (EQ1) in the test period from January 2016 to October 2019.}
\label{fig:strat_ls_vg1}
\end{figure}
\begin{figure}[!htbp]
\centering
\includegraphics[width=\figscale\linewidth]{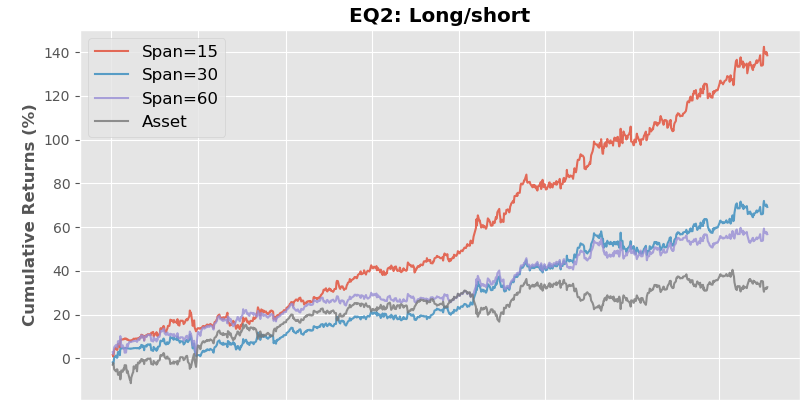} 
\includegraphics[width=\figscale\linewidth]{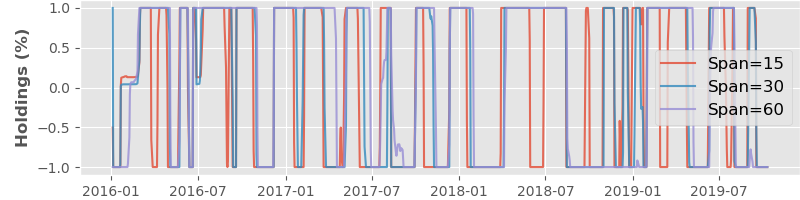}
\caption{Cumulative returns of long/short HMM strategies for instrument $I_{9}$ (EQ2) in the test period from January 2016 to October 2019.}
\label{fig:strat_ls_z_1}
\end{figure}
\begin{figure}[!htbp]
\centering
\includegraphics[width=\figscale\linewidth]{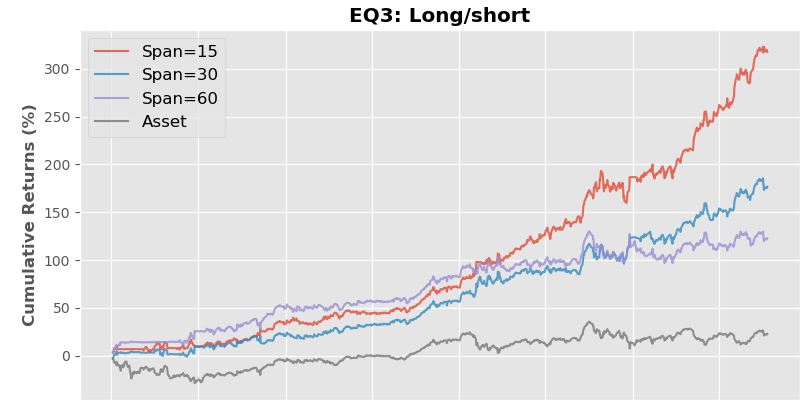} 
\includegraphics[width=\figscale\linewidth]{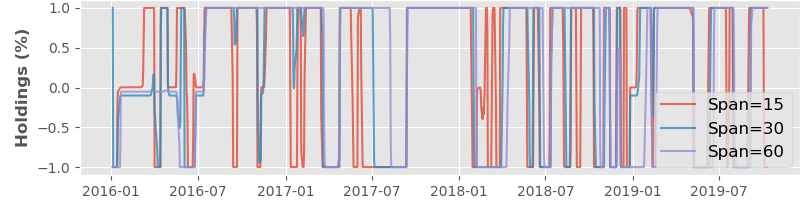}
\caption{Cumulative returns of long/short HMM strategies for instrument $I_{10}$ (EQ3) in the test period from January 2016 to October 2019.}
\label{fig:strat_ls_nk1}
\end{figure}
\begin{figure}[!htbp]
\centering
\includegraphics[width=\figscale\linewidth]{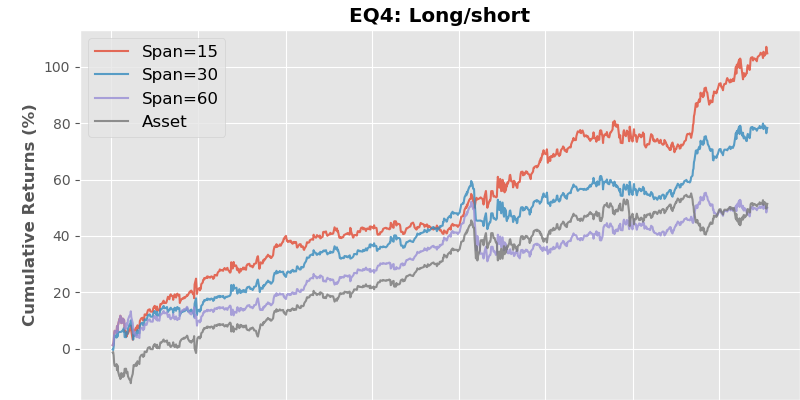} 
\includegraphics[width=\figscale\linewidth]{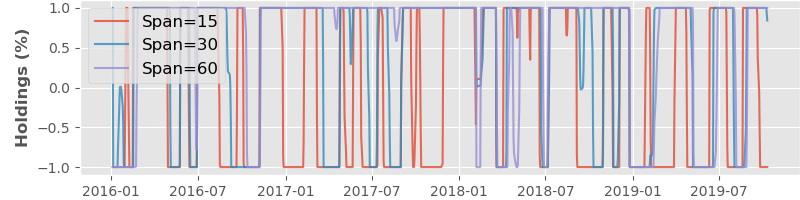}
\caption{Cumulative returns of long/short HMM strategies for instrument $I_{11}$ (EQ4) in the test period from January 2016 to October 2019.}
\label{fig:strat_ls_es1}
\end{figure}
\begin{figure}[!htbp]
\centering
\includegraphics[width=\figscale\linewidth]{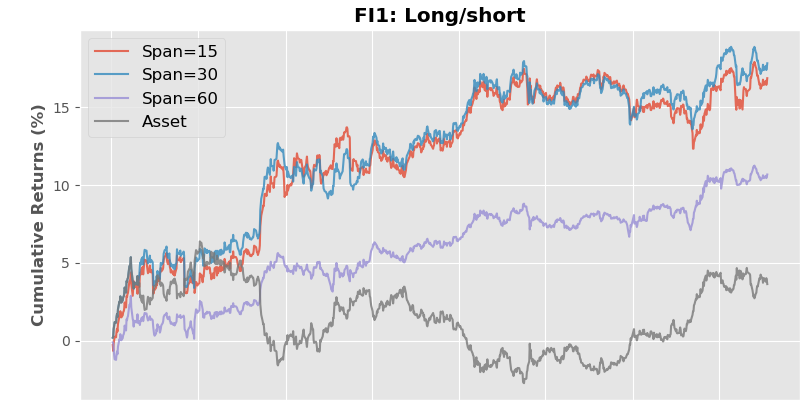} 
\includegraphics[width=\figscale\linewidth]{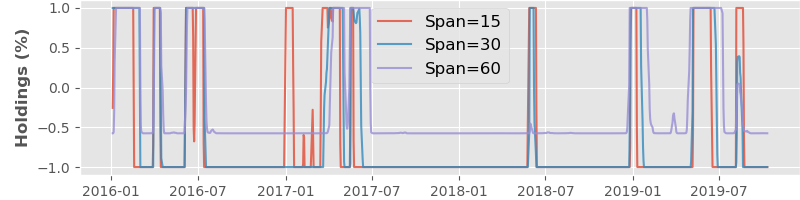}
\caption{Cumulative returns of long/short HMM strategies for instrument $I_{12}$ (FI1) in the test period from January 2016 to October 2019.}
\label{fig:strat_ls_ty1}
\end{figure}
\begin{figure}[!htbp]
\centering
\includegraphics[width=\figscale\linewidth]{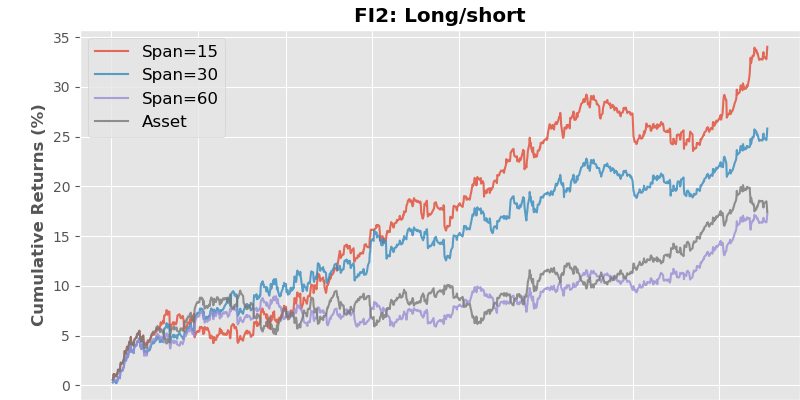} 
\includegraphics[width=\figscale\linewidth]{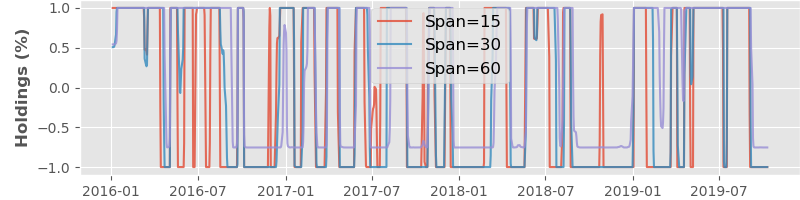}
\caption{Cumulative returns of long/short HMM strategies for instrument $I_{13}$ (FI2) in the test period from January 2016 to October 2019.}
\label{fig:strat_ls_rx1}
\end{figure}
\begin{figure}[!htbp]
\centering
\includegraphics[width=\figscale\linewidth]{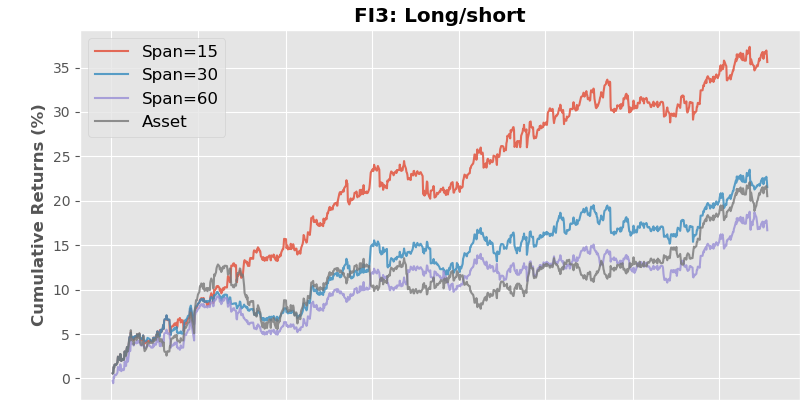} 
\includegraphics[width=\figscale\linewidth]{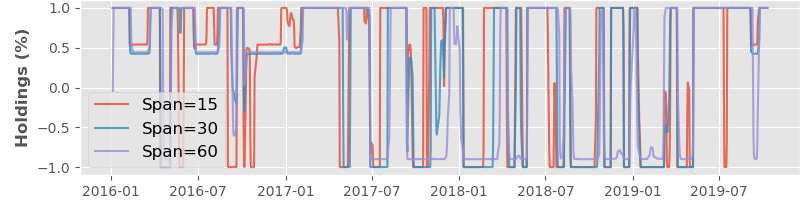}
\caption{Cumulative returns of long/short HMM strategies for instrument $I_{14}$ (FI3) in the test period from January 2016 to October 2019.}
\label{fig:strat_ls_g_1}
\end{figure}
\begin{figure}[!htbp]
\centering
\includegraphics[width=\figscale\linewidth]{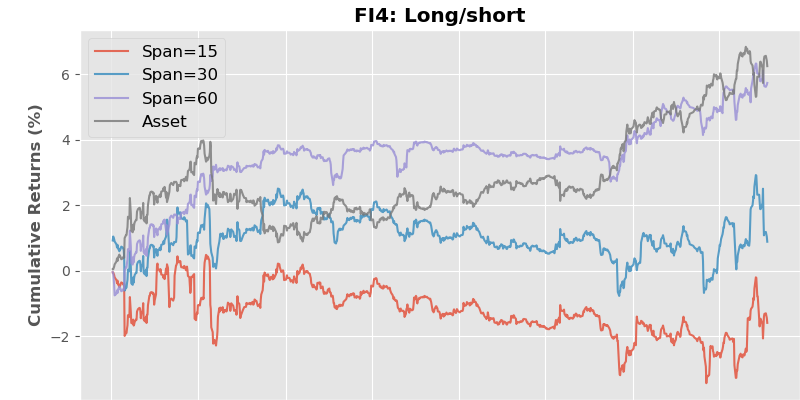} 
\includegraphics[width=\figscale\linewidth]{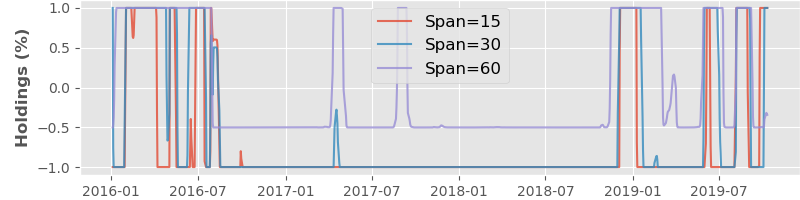}
\caption{Cumulative returns of long/short HMM strategies for instrument $I_{15}$ (FI4) in the test period from January 2016 to October 2019.}
\label{fig:strat_ls_jb1}
\end{figure}

Figure \ref{fig:strat_l_gc1}-\ref{fig:strat_l_jb1} and Figure \ref{fig:strat_ls_gc1}-\ref{fig:strat_ls_jb1} shows the HMM strategies long-only and long/short, respectively, based on the outcomes of $\text{PESR}^{3}_{s}(1)_{s \in \{15, 30, 60\}}$ for the instruments $I$.

\end{document}